\documentclass[a4paper,12pt]{article}
\usepackage{amssymb}

\usepackage{hyperref}
\usepackage{subcaption}
\usepackage{xcolor}
\usepackage{mathrsfs}
\usepackage{graphicx}
\usepackage{bbm}
\usepackage{latexsym}
\usepackage{dsfont}
\usepackage{amsmath}
\usepackage[title,titletoc]{appendix}
\usepackage[super]{nth}
\usepackage{cancel}
\usepackage{booktabs}
\usepackage{tikz}
\usepackage{tikz-feynman} % requires lualatex or xelatex
\tikzfeynmanset{compat=1.1.0}
\usepackage{multicol}
\usepackage{bbold}
\usepackage{amsfonts}
\usepackage{cite}
\usepackage{cleveref}
\usepackage[normalem]{ulem}
\usepackage{comment}
\usepackage{tikz-feynman}
\usepackage{amsmath}

\newcommand{\mathsym}[1]{{}}

\newcommand{\qed}{\nobreak \ifvmode \relax \else \ifdim\lastskip<1.5em \hskip-\lastskip \hskip1.5em plus0em minus0.5em \fi \nobreak \vrule height0.75em width0.5em depth0.25em\fi}
\newcommand{\diag}{\mbox{diag}}

\def\app#1#2{  \mathrel{    \setbox0=\hbox{$#1\sim$}    \setbox2=\hbox{      \rlap{\hbox{$#1\propto$}}      \lower1.1\ht0\box0    }    \raise0.25\ht2\box2  }}

\begin{document}

	\begin{titlepage} 
		\begin{center} \hfill \\
		\hfill \\
			\textbf{\Large
Feasible Deviations from Unitarity with Vector-Like Quark Singlets}

			 \vskip 0.1cm Francisco Albergaria $^{a,}$\footnote{francisco.albergaria@tecnico.ulisboa.pt}, Francisco J. Botella $^{b,}$\footnote{francisco.j.botella@uv.es}, 
             G.C. Branco $^{a,}$\footnote{gbranco@tecnico.ulisboa.pt},\\ 
José Filipe Bastos $^{a,}$\footnote{jose.bastos@tecnico.ulisboa.pt}
and
J.I. Silva-Marcos $^{a,}$\footnote{juca@cftp.tecnico.ulisboa.pt}

\vskip 0.01in $^a$ Centro de
F{\'\i}sica Te\'orica de Part{\'\i}culas, CFTP, \\ Departamento de
F\'{\i}sica,\\ {\it Instituto Superior T\'ecnico, Universidade de Lisboa, }
\\ {\it Avenida Rovisco Pais nr. 1, 1049-001 Lisboa, Portugal}

\vskip 0.01in $^b$ Departament de Física Teòrica and IFIC, \\ {\it Universitat de València-CSIC, E-46100 Burjassot, Spain}  \end{center}

		\begin{abstract} 
We deduce pertinent relations between the elements of the CKM matrix, and find that not all of these are totally compatible with experiment and/or the assumption of the SM $3 \times 3$ unitarity. We identify relations between moduli and complex phases of the CKM-elements which may signal deviations from unitary (DUs).

Recent analyses of $|V_{cs}|$ appear to be in tension with the SM assumption of CKM unitarity in the second row, similar to previous hints of DUs in the first row from previous $|V_{ud,s}|$ results. Here, we explore DUs induced in VLQ iso-singlet models, in particular the possibility of having significant DUs of the first and second rows of the CKM matrix.

We perform a extensive analysis of $n\leq 5$ VLQ iso-singlet models and identify a useful set of parametrizations with the intention of coherently exploring  parameter space. We assess the feasibility of each model and confront it with constraints from key flavor observables, with particular attention to the neutral kaon and $D^0$-meson sectors, especially $\epsilon_K$ and $x_D$.
We find that smaller $n\leq 2$ VLQ-models cannot accommodate the new $|V_{cs}|$ results in the DUs of the second CKM-row. In contrast, larger models extending both up- and down-type sectors, can indeed produce significant second-row DUs, e.g. in a model with 3 up- and 2 down-type VLQs, even reaching the order of the Cabibbo angle. To our knowledge, the phenomenology of extensions combining both up- and down-type VLQs has yet to be thoroughly studied.

Finally, we find that VLQ-singlet models may induce very large enhancements in the rephasing invariant phase $\beta_K$, entering in the Cabibbo sector.
\end{abstract}
	\end{titlepage}
	
\section{Introduction}
	
One of the simplest and best motivated extensions of the Standard Model (SM) consists of adding Vector-Like Quarks (VLQs), i.e. quarks with left- and right-handed components transforming in the same way under the SM gauge group $SU(3)_c\times SU(2)_L\times U(1)_Y$. One finds VLQs in many Beyond the SM (BSM) models such as grand unification models and scenarios with extra dimensions. They also emerge in models addressing the electroweak hierarchy problem in which the light Higgs arises as a pseudo-Goldstone boson of a global symmetry.
In models with inter-family symmetries, VLQ models may possibly contribute to finding an origin for fermion mass hierarchies and mixings. In axion models as well as in axion-less models of the Nelson-Barr type they may provide solutions to the strong CP problem. An extensive list of all these BSM models can be found in \cite{Albergaria:2024pji}.

All in all, VLQ models play an important role in attempts at solving some of the open questions in particle physics. Signatures of VLQs may appear in future experiments, or show up in further analyses of existing LHC data. Potential imprints of mixing with VLQs also arise from evidence of the CKM unitarity violation. Such effects may manifest themselves as deviations from the unitarity (DUs) conditions of the CKM rows or columns, or as discrepancies with respect to the SM expectations for unitarity triangle angles and other rephasing-invariant CP-violating phases. Moreover, in models with iso-singlet VLQs, DUs in the rows are directly connected to the characteristic decay patterns expected for the heavy quarks. For instance, the measurement of large DUs predominantly in the first row, could signal dominant coupling of a vector-like $T$ quark to $u$ and $d$ quarks, compared to suppressed couplings to $t$ and $b$ quarks in channels involving the $W$, $Z$, or Higgs bosons \cite{Botella:2021uxz}. In turn, each of the possible VLQ decay patterns will also set different bounds on the expected mass of these BSM fields.

In this work we specifically focus on the open question of potential deviations from unitarity of the quark mixing parameterized by the Cabibbo Kobayashi Maskawa (CKM) matrix ($V_{\text{CKM}}$). 
DUs occur when the mixing can no longer be represented by a $3\times 3$ unitary matrix. Indeed, recent results point to significant DUs in the first row of $V_{\text{CKM}}$ \cite{Belfatto:2021jhf,Belfatto:2023tbv,Crivellin:2022rhw,Cirigliano:2023nol,Crivellin:2021bkd,Crivellin:2020lzu,Cirigliano:2022yyo,Cheung:2020vqm} of the order $\textstyle\tiny{{\sqrt{1-\sum_{j=1}^3{|V_{uj}|^2}}}}  \sim 0.04 $. 

Additionally, a
recent analysis of rare charm to strange decays \cite{Bolognani:2024cmr} obtains for the matrix element $|V_{cs}|=0.957\pm 0.003$. This is smaller than the PDG value. In fact, assuming the PDG input for $|V_{cd}|$ and $|V_{cb}|$, this new result for $|V_{cs}|$ would imply
a surprisingly large deviation of unitarity in the second row of the mixing, almost up to the order of the Cabibbo angle, in stark contrast to the SM prediction of CKM unitarity.

In view of this remarkable result, the main goal of this work is not only to explain DUs of the first row, but above all, it is an attempt to explain the potential DUs in the second row as suggested by \cite{Bolognani:2024cmr}, together with DUs in the columns (and subsequent tensions which herein arise). We shall attempt this with the simplest extension of the SM introducing  VLQ-singlets, i.e. VLQs with quantum numbers $(\mathbf{3},\mathbf{1})_{2/3}$ for up-type fields and $(\mathbf{3},\mathbf{1})_{-1/3}$ for down-type fields.  
In particular, we are interested in setting limits, coming from flavour observables, on the deviations from unitarity of the new physics induced by these VLQ-singlets, and in finding which models might allow for unexpected large DUs in the second row. 

In the next chapter, and starting from the assumption that the CKM matrix is a $3 \times 3$ unitary matrix, we deduce pertinent relations between its elements, which include the moduli and/or its rephasing invariant phases. We identify stress-tests and signals which could point to potential DUs, and confront these with experiment and the assumption of the $3 \times 3$ unitarity.

In Chapter 3 and 4, we perform a comprehensive analysis of the parameter space of VLQ models with a low number of singlets up to two VLQs. We further test the feasibility of each model, including valuable restrictions on a series of the most relevant quantities, as, for instance, the most prominent mixing matrix elements, and the usual unitary triangle angles. Besides these, special attention is given to important flavor observables, e.g. $x_D$ and $\epsilon_K$, as well as to several other phenomenologically quantities that may significantly constraint the parameter space of these models. All these are inserted into a $\chi^2$ analysis. Another important aspect of this study is the identification of a suitable parametrization that allows the parameter space of each model to be explored in a consistent and systematic manner.

In Chapter 4 specifically, we explore the three VLQ scenarios involving two new VLQs. We find that in all these models (be it either only two up-types, or only two down-types, or a combination of one up- and one down-type VLQs) the deviations from unitarity of the second row roughly accompany the DUs of the first row and cannot achieve the values/results suggested by \cite{Bolognani:2024cmr}.

In Chapter 5, with much more success (and in view of the analysis in \cite{Bolognani:2024cmr}), we are able to obtain very significant deviations from unitary in the second row using more elaborate and complex models with larger combinations of up-type and down-type VLQs. This is not only possible due the unique possibilities offered by a much larger parameter-space, but also due to the mingling effect of the two quark-type sectors of VLQs, implying the inclusion of new contributions and extra terms in the flavour observables. The latter were, until now, overlooked in the literature.

Finally, in Chapter 6, we present our conclusions.

\section{Model-Independent Approaches to Deviations from Unitarity of the CKM Matrix}

In the following, we test the unitarity of the CKM mixing.
Starting from the assumption that the CKM matrix is a $3 \times 3$ unitary matrix, and then generalizing further to incorporate larger matrices, we obtain diverse and conclusive relations between its elements. These relations may be between the moduli of the CKM elements only, or include their complex phases. As we shall see, not all of these are totally compatible with experiments and/or the assumption of $3 \times 3$ unitarity.

The issue of deviations from unitarity in the rows of the CKM mixing matrix has been previously approached in the literature \cite{Alves:2023ufm}. In particular, the viability of DUs in the first row has been tested in \cite{Branco:2021vhs,Botella:2021uxz}. However, DUs may also occur in the columns.

\subsection{Deviations from Unitarity in Columns and Rows}
\begin{figure*}
\centering
\resizebox{0.75\textwidth}{!}{%
  \includegraphics[scale=0.6]{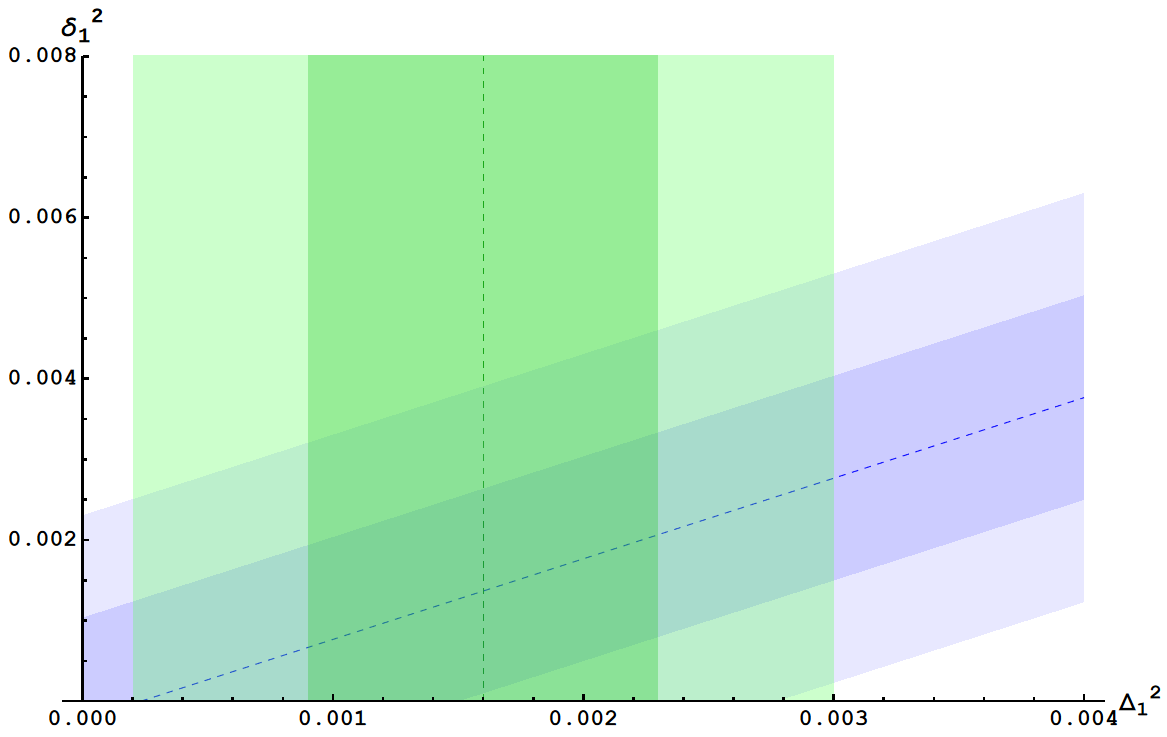}
}
\caption{Deviations from unitarity of the first column of the CKM matrix squared ($\delta^2_1$) \textit{vs} deviations from unitarity of the first row of the CKM matrix squared ($\Delta^2_1$). The dark(light) green regions represents the allowed regions at $1 \sigma$ ($2 \sigma$) for the deviation from unitarity of the first row of the CKM matrix $\left(\Delta^\text{exp}_1\right)^2=(1.6\pm0.7)\times 10^{-3}$ \cite{ParticleDataGroup:2024cfk}, corresponding to the central value $\Delta^\text{exp}_1\approx 0.04$. The dark(light) blue regions represent the regions obtained from \cref{eq:delta1Delta1} at $1 \sigma$ ($2 \sigma$), considering the PDG values for $|V_{td}|$, $|V_{us}|$, $|V_{cd}|$, $|V_{ub}|$. }	
\label{fig:deltavsDelta}       
\end{figure*}

Let us suppose for now that besides a deviation from unitarity in the first row of the CKM matrix (as suggested by the results of \cite{flavorLatticeAveragingGroupFLAG:2024oxs}), there is also a DU in the first column. We then find that these deviations from unitarity are related. When having both of these deviations, one necessarily has that
\begin{subequations}
\allowdisplaybreaks
\begin{align}
    |V_{ud}|^2 + |V_{us}|^2 + |V_{ub}|^2 &= 1 - \Delta_1^2,
    \\
    |V_{ud}|^2 + |V_{cd}|^2 + |V_{td}|^2 &= 1 - \delta_1^2,
\end{align}
\end{subequations}
with $\Delta_1$ the DU of the first row and $\delta_1$ the DU of the first column. Subtracting and rearranging the two equations, we find 
\begin{equation} \label{eq:delta1Delta1}
    \delta_1^2 = \Delta_1^2 - |V_{td}|^2 + |V_{us}|^2 - |V_{cd}|^2 + |V_{ub}|^2.
\end{equation}
In Fig.~\ref{fig:deltavsDelta}, we plot this relation thereby visualizing the deviations from unitarity of the first column of the CKM matrix as a function of the deviations from unitarity of the first row. The PDG  values \cite{ParticleDataGroup:2024cfk} were used for the mixing moduli $|V_{td}|$, $|V_{us}|$, $|V_{cd}|$ and $|V_{ub}|$, with their corresponding variations, as well as the DU of the first row
\begin{equation}\label{eq:Delta_1}
    \left(\Delta^\text{exp}_1\right)^2=(1.6\pm0.7)\times 10^{-3}.
\end{equation}
From this plot one concludes that the deviations from unitarity of the first column tend to be of the same order, but may be slightly larger than the deviations from unitarity of the first row.

\subsubsection*{Second Row}

Apart from the DUs in the first row of the CKM matrix, salient new results from \cite{Bolognani:2024cmr} appear to suggest a possible deviation from the unitarity condition of the second row of the CKM. The analysis performed on leptonic and semi-leptonic decays $c \to s l^+ \nu$ using recent BESIII data, resulted in $|V_{cs}|_\text{new} = 0.957 \pm 0.003$
which is smaller than the corresponding PDG value.

Furthermore, using this value for $|V_{cs}|$, along with the most recent PDG values for $|V_{cd}|$ and $|V_{cs}|$, we obtain
the following result
\begin{equation}\label{eq:D2_squared}
    \Delta^2_2=\left(3.20\pm 0.58\right)\times 10^{-2},
\end{equation}
for the DU in the second row of the CKM matrix, here defined as
\begin{equation}
    \Delta^2_2\equiv 1- |V_{cd}|^2 - |V_{cs}|^2 - |V_{cb}|^2.
\end{equation}
This seems to point to a tension of around $5.5\sigma$ from the SM assumption of $3\times 3$ unitarity.
The resulting DU is also very large, being of the order of the Cabibbo angle,
\begin{equation}\label{eq:Delta_2}
\Delta_2=0.179\pm 0.016.
\end{equation}
One (if not the) major goal of this work is to try to accommodate these striking results. 

We also note that, apart from the theoretical analysis in \cite{Bolognani:2024cmr} which presents a particularly small value for $|V_{cs}|$ when compared with the the PDG global average, some recent experimental results have obtained values for $|V_{cs}|$ that trend in that same direction. Particularly, determinations coming from semi-leptonic $D$ decays have, over the past few years, contributed to the reduction of this $|V_{cs}|$ average value. For instance, consider the results coming from the BESIII Collaboration \cite{BESIII:2024slx,BESIII:2026uin}, which have obtained $|V_{cs}|=0.962\pm 0.003$ and $|V_{cs}|=0.9639\pm 0.0045$, respectively.

\subsection{Deviations from Unitarity in the CKM columns}

We further analyse DUs which may occur in the columns.
Assuming that the CKM matrix is indeed a $3 \times 3$ unitary matrix, this implies that the sums of the moduli-squared of the elements of each row and column are equal to $1$. 

\subsubsection*{First Column}

For the case of the first column, we have
\begin{equation}
    |V_{ud}|^2 + |V_{cd}|^2 + |V_{td}|^2 = 1.
\end{equation}
Rearranging this equation, we write $|V_{td}|$ in terms of $|V_{ud}|$ and $|V_{cd}|$ as
\begin{equation}
\label{eq:1stcolumn2}
    |V_{td}|^2 = 1 - |V_{ud}|^2 - |V_{cd}|^2.
\end{equation}
Thus, with the knowledge of the experimental values of $|V_{ud}|$ and $|V_{cd}|$ we can compute $|V_{td}|$, explicitly assuming that the CKM matrix is a $3 \times 3$ unitary matrix \footnote{We compute $|V_{td}|$ in terms of $|V_{ud}|$ and $|V_{cd}|$ because $|V_{ud}|$ and $|V_{cd}|$ are measured at tree level, while the value that we have for $|V_{td}|$ results from a loop-level measurement. Thus, one expects the measurements of $|V_{ud}|$ and $|V_{cd}|$ to be more precise than that of $|V_{td}|$. Indeed, we find from the PDG~\cite{ParticleDataGroup:2024cfk} that $|V_{ud}|$ and $|V_{cd}|$ have smaller relative uncertainties than $|V_{td}|$.}. For this purpose, we use the PDG input values for $|V_{ud}|$ and $|V_{cd}|$, which are~\cite{ParticleDataGroup:2024cfk}
\begin{subequations}
    \allowdisplaybreaks
\begin{align}
    |V_{ud}| &= 0.97367 \pm 0.00032
    \\
    |V_{cd}| &= 0.2247 \pm 0.0022.
\end{align}
\end{subequations}
Inserting these experimental values in the previous relations, we get
\begin{equation}
    |V_{td}|_{\text{unit}} = 0.038 \pm 0.015.
\end{equation}
However, the experimental value that one actually finds in the PDG for $|V_{td}|$ is $|V_{td}|_{\text{PDG}} = 0.0086 \pm 0.0002$. This means that the value for $|V_{td}|$ computed from the experimental values given by the PDG for $|V_{ud}|$ and $|V_{cd}|$ assuming unitarity of the first column of the CKM matrix ($|V_{td}|_{\text{unit}}$) and the value for $|V_{td}|$ from the PDG ($|V_{td}|_{\text{PDG}}$) are $1.96 \sigma$ away from each other. All in all, this disagreement points towards a deviation from unitarity in the first column of the CKM matrix.

Still, we can go further in the supposed framework of $3 \times 3$ unitarity. This is because $|V_{td}|$ can also be obtained from another $3\times 3$ unitarity relation,
\begin{equation}
    |V_{td}|^2 = |V_{us}|^2 - |V_{cd}|^2 + |V_{ub}|^2,
\end{equation}
without having to take into account the CKM element $V_{ud}$. 
Then, using the PDG values,
\begin{subequations}
\allowdisplaybreaks
\begin{align}
    |V_{us}| =& \ 0.22431 \pm 0.00085,
    \\
    |V_{cd}| =& \ 0.2247 \pm 0.0022,
    \\
    |V_{ub}| =& \left( 3.89 \pm 0.16 \right) \times 10^{-3},
\end{align}
\end{subequations}
we get
\begin{equation}
    |V_{td}|^2 = -0.00016 \pm 0.00106.
\end{equation}
However, this is perfectly compatible with the PDG value $|V_{td}|^2_{\text{PDG}} = \left(7.4 \pm 0.3\right) \times 10^{-5}$, being the central values at $0.22 \sigma$.

From all this, we may conclude that the values of $|V_{us}|$ and $|V_{ub}|$ appear to be more consistent with $3 \times 3$ unitarity than the analysis where one invokes the value of $|V_{ud}|$ (as input), thus reinforcing the so-called Cabibbo Angle Anomaly (CAA).

\subsubsection*{Second Column}
A similar reasoning can be followed for the case of the second column of the CKM matrix, where, in case of unitarity, we find 
\begin{equation} \label{eq:unit_2column}
    |V_{ts}|^2 = 1 - |V_{us}|^2 - |V_{cs}|^2.
\end{equation}
which with the PDG input values~\cite{ParticleDataGroup:2024cfk} for $|V_{us}|$ and $|V_{cs}|$
\begin{subequations}
    \allowdisplaybreaks
\begin{align}
    |V_{us}| &= 0.22431 \pm 0.00085,
    \\
    |V_{cs}| &= 0.969 \pm 0.005,
\end{align}
    \label{vusvcs}
\end{subequations}
results in
\begin{equation}
    |V_{ts}|_{\text{unit}}^2 = 0.011 \pm 0.010.
\end{equation}
The experimental value that one finds in the PDG for $|V_{ts}|$ is, however, $|V_{ts}|_{\text{PDG}} = \left(41.5 \pm 0.9 \right) \times 10^{-3}$, which corresponds to $|V_{ts}|^2_{\text{PDG}} = \left(1.72 \pm 0.07 \right) \times 10^{-3}$. Thus, $|V_{ts}|_{\text{unit}}^2$ and $|V_{ts}|^2_{\text{PDG}}$ are $0.93 \sigma$ away from each other, which means that the experimental results (until now) referred in the PDG are in agreement with the $3\times3$ unitarity-condition of the second column. 

However, using the new value of $|V_{cs}| = 0.957 \pm 0.003$ from \cite{Bolognani:2024cmr}, in the unitarity expression from \cref{eq:unit_2column}, we get
\begin{equation}
    |V_{ts}|_{\text{unit, new}}^2 = 0.034 \pm 0.006.
\end{equation}
Thus, with the new input, we find is that $|V_{ts}|_{\text{unit, new}}$ and $|V_{ts}|_{\text{PDG}}$ are now $5.58 \sigma$ away from each other. Therefore, considering the value for $|V_{cs}|$ from \cite{Bolognani:2024cmr}, we get a strong disagreement between the value for $|V_{ts}|$ computed using unitarity and the value for $|V_{ts}|$ from the PDG, which points to a deviation from unitarity, affecting also the second column of the CKM matrix.

If, on the other hand, we used the $3 \times 3$ unitarity relation, not involving the matrix-element $|V_{cs}|$, where
\begin{equation}
    |V_{ts}|^2 = |V_{cd}|^2 - |V_{us}|^2 + |V_{cb}|^2.
\end{equation}
we would obtain with the input from the PDG
\begin{subequations}
\allowdisplaybreaks
\begin{align}
    |V_{us}| =& \ 0.22431 \pm 0.00085,
    \\
    |V_{cd}| =& \ 0.2247 \pm 0.0022,
    \\
    |V_{cb}| =& \left( 40.7 \pm 1.3 \right) \times 10^{-3},
\end{align}
\end{subequations}
the value
\begin{equation}
    |V_{ts}|^2 = 0.0018 \pm 0.0011.
\end{equation}
This is compatible at less than $1 \sigma$ with the PDG value. Again, this clearly reinforces the fact that a possible new value of $|V_{cs}|$, such as the one obtained in ~\cite{Bolognani:2024cmr},  generates two important problems of unitarity both in the second row and the second column.

Finally, one can look at $|V_{tb}|$ involving the supposed $3\times 3$ unitarity conditions. If one expresses $|V_{tb}|$ as in $|V_{tb}|^2 = 1 - |V_{ub}|^2 - |V_{cb}|^2$, then one finds that $|V_{tb}|$ is compatible with the PDG value. However, when one considers
\begin{subequations}
\allowdisplaybreaks
\begin{align}
    |V_{tb}|^2 =& \ |V_{ud}|^2 + |V_{us}|^2 - |V_{cb}|^2,
    \\
    |V_{tb}|^2 =& \ |V_{cd}|^2 + |V_{cs}|^2 - |V_{ub}|^2.
\end{align}
\end{subequations}
and test these using the recent $|V_{ud}|$ values and the value $|V_{cs}|_{\text{new}}$ obtained in \cite{Bolognani:2024cmr}, then
from the first, one obtains
\begin{equation}
    |V_{tb}|^2 = 0.99669 \pm 0.00074,
\end{equation}
which is at $2.01 \sigma$ from the PDG value $|V_{tb}|^2_{\text{PDG}} = 0.998231^{+0.000058}_{-0.000068}$. From the second relation, we get
\begin{equation}
    |V_{tb}|^2 = 0.966 \pm 0.006,
\end{equation}
which is at $5.37 \sigma$ from the PDG. 

These two estimates, caused by potential deviations from unitarity in the first
and second rows and columns, thus represent possible new correlated influences on the values of the matrix-element $|V_{tb}|$.

\subsection{DUs and the CKM Complex Phases}
\label{ckmphase}
So far, we have focused on the tensions between the moduli of the CKM elements hinting at the fact that the CKM matrix may not be a unitary matrix.

One obvious possibility for non-$3\times 3$-unitary is that the known $3 \times 3$ CKM matrix is part of a larger unitary matrix. The simplest hypothesis is that it is the upper-left block of a larger $4 \times 4$ unitary matrix. This is the case in models with one vector-like singlet (either up- or down-type).
In these models, the $4 \times 4$ unitary matrix, in which the $3 \times 3$ CKM matrix is embedded, has 9 non-factorizable phases. To parameterize the usual set of four of these phases, we use the known rephasing invariants, capturing these phases, from the CKM matrix elements and which are given by
\begin{subequations}
\allowdisplaybreaks
\begin{align}
    \gamma \equiv& \ \mathrm{arg} \left(- V_{ud} V_{cb} V_{ub}^*  V_{cd}^* \right),
    \\
    \beta \equiv& \ \mathrm{arg} \left(- V_{cd} V_{tb} V_{cb}^*  V_{td}^* \right),
    \\
    \beta_s \equiv& \ \mathrm{arg} \left(- V_{cb} V_{ts} V_{cs}^*  V_{tb}^* \right),
    \\
    \beta_K \equiv& \ \mathrm{arg} \left(- V_{us} V_{cd} V_{ud}^*  V_{cs}^* \right).
\end{align}
\label{angles}
\end{subequations}
The phases $\beta_s$ and $\beta_K$ are sometimes referred to as $\chi$ and $\chi^\prime$, respectively. In \cite{Botella:2002fr}, a set of exact relations between the moduli of the CKM matrix elements and the rephasing invariants $\gamma$, $\beta$, $\beta_s$ and $\beta_K$, resulting from $3 \times 3$ unitarity of the CKM matrix, were derived. 

However, the known relations between the moduli and complex phases of the CKM matrix elements have to be modified when the CKM matrix is not a $3 \times 3$ unitary matrix, but, instead, part of a larger $4 \times 4$ unitary matrix. To parameterize the other five phases, we may choose the following rephasing invariants
\begin{subequations}
\allowdisplaybreaks
\begin{align}
    \xi_1 \equiv& \ \mathrm{arg} \left(- V_{14} V_{cd} V_{ud}^*  V_{24}^* \right),
    \\
    \xi_2 \equiv& \ \mathrm{arg} \left( V_{cb} V_{34} V_{tb}^*  V_{24}^* \right),
    \\
    \xi_3 \equiv& \ \mathrm{arg} \left(- V_{cb} V_{41} V_{cd}^*  V_{43}^* \right),
    \\
    \xi_4 \equiv& \ \mathrm{arg} \left( V_{cb} V_{42} V_{cs}^*  V_{43}^* \right),
    \\
    \xi_5 \equiv& \ \mathrm{arg} \left( V_{cb} V_{44} V_{24}^*  V_{43}^* \right).
\end{align}
\end{subequations}
With these nine phases, we obtain the following phase convention
\begin{equation}
    \mathrm{arg} (V) = \begin{pmatrix}
    0 & \beta_K & - \gamma & \xi_1
    \\
    \pi & 0 & 0 & 0
    \\
    -\beta & \pi + \beta_s & 0 & \xi_2
    \\
    \xi_3 & \xi_4 & 0 & \xi_5
\end{pmatrix},
\label{betagamma}
\end{equation}
where the phases in the upper-left $3 \times 3$ block are the usual ones.

From the imaginary part of the orthogonality relation between the first two rows, we then find e.g.
\begin{equation}
    \sin \beta_K = \frac{|V_{ub}| |V_{cb}|}{|V_{us}| |V_{cs}|} \sin \gamma - \frac{|V_{14}| |V_{24}|}{|V_{us}| |V_{cs}|} \sin \xi_1. \label{eq:generalized8}
\end{equation}
If indeed CKM mixing is $3 \times 3$ and unitary then, the term proportional to $\sin \xi_1$ is absent and from the PDG values for the CKM matrix moduli and $\sin \gamma$, we should obtain for $\sin \beta_K = \left(6.67 \pm 0.38 \right) \times 10^{-4}$ \cite{Aguilar-Saavedra:2004roc}.

If however $\sin \beta_K$ is measured to be much larger than this, then, also from this relation, we obtain a clear hint for the existence of DUs.

Some phases, measured in charged current processes, can be tree level observables at leading order, as e.g. $\gamma$, or $\beta_K$ which
might be a tree level observable not being related to matrix elements involving
the top quark. In particular, because the decays $D^0\rightarrow \pi^+ \pi^-, K^+K^-$ at tree
level are controlled by $V^*_{cd} V_{ud}$ and $V^*_{cs} V_{us}$ respectively, it is natural to attribute 
the difference of CP asymmetries $\Delta A_{CP}= A_{CP}(K^+K^-)-A_{CP}(\pi^+ \pi^-) =(-0.154\pm0.029)\% $
to the phase
$\beta_K= \mathrm{arg}(-(V^*_{cs} V_{us})(V^*_{cd} V_{ud}))$. In fact, in the analysis of \cite{Bediaga:2022sxw} the final state interaction needed in these asymmetries is introduced by the strong scattering $\pi^+ \pi^-\rightarrow K^+K^-$ and, roughly speaking, with a value of $\beta_K$ which is double of the SM value, it is easier to accommodate the LHCb measured asymmetry $\Delta A_{CP}$. It is in this framework that we refer to the second term in the rhs of \cref{eq:generalized8}, violating $ 3\times 3 $ unitarity, which can increase the value of $\beta_K$. 

\begin{figure*}[t]
\centering
\resizebox{0.75\textwidth}{!}{%
  \includegraphics[scale=0.1]{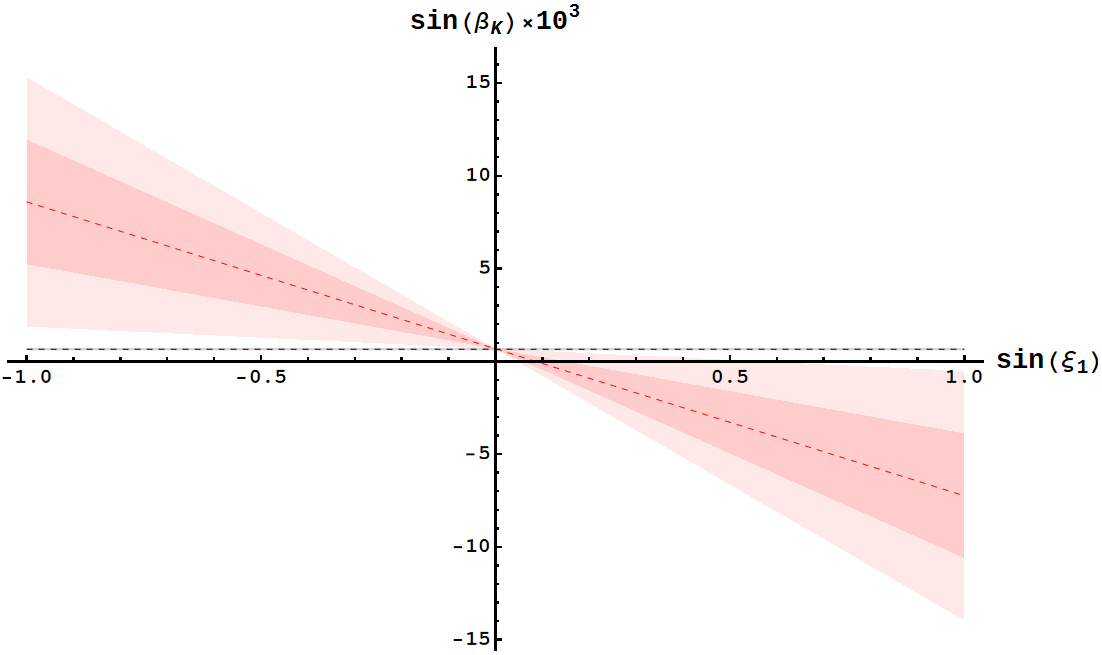}
}
\caption{Plot of $\sin \beta_k$ as a function of $\sin \xi_1$ as in \cref{eq:generalized8}.
The horizontal dashed line represents the value of $\sin\beta_K$ in the SM. We use the PDG values for the $\gamma$ phase and the moduli of the CKM matrix elements without assuming unitarity. In red we present the 1$\sigma$ (darker) and 2$\sigma$ (lighter) regions for $\sin\beta_K$ allowing for deviations of unitarity in the first and second row. Here we use $|V_{14} ||V_{24}|=\left(\Delta^\text{exp}_1\right)^2=(1.6\pm0.7)\times 10^{-3}$.}
\label{fig:betaK_xi1}       
\end{figure*}

Such considerations and the type of models which induce these effects will be studied here.
To form a picture of DUs inspired by the relations obtained from the mixing complex phases, we plot, in Fig. \ref{fig:betaK_xi1}, $\sin \beta_K $ against a possible new phase $\sin \xi_1 $ coming from a larger matrix.

With regard to $\sin \beta_s $, a similar relation as \cref{eq:generalized8} is obtained from the second and third row:

\begin{equation}
   \sin \beta_s = \frac{|V_{cd}| |V_{td}|}{|V_{cs}| |V_{ts}|} \sin \beta + \frac{|V_{24}| |V_{34}|}{|V_{cs}| |V_{ts}|} \sin \xi_2.
    \label{eq:generalized10}
\end{equation}
If again the CKM mixing is $3 \times 3$ and unitary then, the term proportional to $\sin \xi_2$ is absent and from the PDG values for the CKM matrix moduli and $\sin \beta$, we should obtain\footnote{In fact, the CKMfitter PDG result gives a value of $\sin \beta_s = 0.0184 \pm 0.0004$ assuming no penguins.} for $\sin \beta_s = 0.018 \pm 0.001$.

The $\beta_s$ phase enters, in the SM, at 1 loop level, essentially in the box diagram
controlling $B_s^0 - \bar{B}_s^0$ mixing. Therefore, it would not be trivial to assess any deviation from unitarity in the $\beta_s$ value. In a New Physics model at 1 loop, there will be other intrinsic new contributions in addition to the potential deviation from unitarity. Nevertheless, there is some room for these deviations although not too much. The actual measured value is $\sin \beta_s = 0.025\pm0.010$, still fully compatible with the SM value.

\subsubsection*{Interdependent Phase Relations}
More significant relations, which must be in agreement with the previous ones and may test deviations from unitarity, are 
\begin{subequations}
\allowdisplaybreaks
\begin{align} 
    & \sin \beta_K = \frac{|V_{td}| |V_{ts}|}{|V_{ud}| |V_{us}|} \sin \left( \beta + \beta_s \right) + \frac{|V_{41}| |V_{42}|}{|V_{ud}| |V_{us}|} \sin \left( \xi_3 - \xi_4 \right),
    \\
    \label{eq:generalized12}
    & \sin \beta_s = \frac{|V_{us}| |V_{ub}|}{|V_{ts}| |V_{tb}|} \sin \left( \beta_K + \gamma \right) + \frac{|V_{42}| |V_{43}|}{|V_{ts}| |V_{tb}|} \sin \xi_4.
    \\
   & \frac{|V_{ub}|}{|V_{td}|} = \frac{|V_{tb}|}{|V_{ud}|} \frac{\sin \beta}{\sin \gamma} - \frac{|V_{41}| |V_{43}|}{|V_{td}| |V_{ud}|} \frac{\sin \xi_3}{\sin \gamma}\\
        & |V_{ud}| |V_{td}| \sin \beta - |V_{us}| |V_{ts}| \sin\left( \beta_K - \beta_s \right) - |V_{ub}| |V_{tb}| \sin \gamma + |V_{14}| |V_{34}| \sin \left(\xi_1 - \xi_2 \right) = 0,
\end{align}
\end{subequations}
In addition, from orthogonality between the columns of $V$ by multiplying the orthogonality relations by some phase before taking the imaginary parts, we find
\begin{subequations}
\allowdisplaybreaks
\begin{align}
    \label{eq:generalized13}
    & |V_{ub}| = \frac{|V_{cd}| |V_{cb}|}{|V_{ud}|} \frac{\sin \beta}{\sin \left(\gamma + \beta\right)} - \frac{|V_{41}| |V_{43}|}{|V_{ud}|} \frac{\sin \left( \beta + \xi_3 \right)}{\sin \left(\gamma + \beta\right)},
    \\
    \label{eq:Vtd}
    & |V_{td}| = \frac{|V_{cd}| |V_{cb}|}{|V_{tb}|} \frac{\sin \gamma}{\sin \left(\gamma + \beta\right)} - \frac{|V_{41}| |V_{43}|}{|V_{tb}|} \frac{\sin \left( \gamma - \xi_3 \right)}{\sin \left(\gamma + \beta\right)},
    \\
    & \sin \beta_s = \frac{|V_{us}| |V_{ub}|}{|V_{cs}| |V_{cb}|} \sin \left(- \beta_s + \beta_K + \gamma \right) - \frac{|V_{42}| |V_{43}|}{|V_{cs}| |V_{cb}|} \sin \left( \beta_s - \xi_4 \right).
\end{align}
\end{subequations}
Dividing eq.~\eqref{eq:Vtd} by $|V_{ts}|$ and using normalization of rows and columns, one obtains
\begin{equation}
\begin{split}
    r =& \ \frac{\sin \gamma}{\sin \left(\gamma + \beta\right)} \frac{|V_{cd}|}{|V_{tb}|} \sqrt{1 + r^2 - r^2 \left(\frac{|V_{tb}|}{|V_{ud}|} \frac{\sin \beta}{\sin \gamma} - \frac{|V_{41}| |V_{43}|}{|V_{td}| |V_{ud}|} \frac{\sin \xi_3}{\sin \gamma} \right)^2 + r^2 \frac{|V_{34}|^2 - |V_{43}|^2}{|V_{td}|^2}}
    \\
    & - \frac{|V_{41}| |V_{43}|}{|V_{tb}| |V_{ts}|} \frac{\sin \left( \gamma - \xi_3 \right)}{\sin \left(\gamma + \beta\right)},
\end{split}
\end{equation}
where $r \equiv |V_{td}|/|V_{ts}|$. This is to be compared with a similar relation found in \cite{Botella:2002fr}.

Using eqs.~\eqref{eq:generalized8} and~\eqref{eq:generalized10}, together with normalization conditions, we get
\begin{equation}
\begin{split}
    \sin (\beta_s - \beta_K) =& \ r \sin \beta \frac{|V_{us}|}{|V_{ud}|} \left(1 - \frac{|V_{cb}|^2}{|V_{us}|^2} + \frac{|V_{14}|^2 - |V_{43}|^2}{|V_{us}|^2}  \right. 
    \\
    & \left. - \frac{|V_{tb}||V_{41}||V_{43}|}{|V_{us}|^2 |V_{td}|} \frac{\sin \xi_3}{\sin \beta} - \frac{|V_{ud}||V_{14}||V_{34}|}{|V_{us}|^2|V_{td}|} \frac{\sin \left(\xi_1 - \xi_2 \right)}{\sin \beta} \right).
\end{split}
\end{equation}
Combining Eqs.~\eqref{eq:generalized12} and~\eqref{eq:generalized13}, we get 
\begin{equation}
    \begin{split}
    \sin \beta_s =& \  \frac{|V_{us}||V_{cd}| |V_{cb}|}{|V_{ts}| |V_{tb}||V_{ud}|} \frac{\sin \beta \sin \left( \beta_K + \gamma \right)}{\sin \left(\gamma + \beta\right)}    - \frac{|V_{us}||V_{41}| |V_{43}|}{|V_{ts}| |V_{tb}| |V_{ud}|} \frac{\sin \left( \beta + \xi_3 \right) \sin \left( \beta_K + \gamma \right)}{\sin \left(\gamma + \beta\right)}  
    \\
    &+ \frac{|V_{42}| |V_{43}|}{|V_{ts}| |V_{tb}|} \sin \xi_4.
\end{split}
\label{morebetas}
\end{equation}

A crucial comment on the unitarity relations derived here:
it is clear, in the context of a larger picture of unitarity as is the case with VLQ-singlets, that all the different relations (e.g. from \cref{eq:generalized8} to \cref{morebetas}, and expressing the DUs of the first three rows and columns) must be fulfilled. If however these do not match with experiment or are inconsistent, then, we must seek solutions in other contexts.

\subsubsection*{Unitarity in Larger Matrices} 
When the mixing matrix is not a $4\times 4$ unitary matrix anymore, but a larger one, the picture that is presented in \cref{betagamma} continues, in a certain sense, to be very useful. 

We shall exemplify this for a larger case when the CKM mixing matrix is part of a $5\times 5$ unitary matrix $\mathcal{V}$. To do this, we focus on the relations derived from the orthogonality of the rows. Further generalization for yet larger cases $(4+n)\times (4+n)$, $n>1$ will be evident.

For a $5\times 5$ unitary matrix $\mathcal{V}$, the relations we obtain from the orthogonality of the rows, are derived from the fact that
\begin{equation}
\mathcal{V}\cdot \mathcal{V}^\dagger=\mathbb{1}.
\label{orthos}
\end{equation}

With respect to the first row, we can capture the whole content of the influence of the extra elements $(V_{14}, \, V_{15}) $ of $\mathcal{V}$ on the DU (of the $3\times 3$ part), even if we rearrange $\mathcal{V}$, without affecting the orthogonality condition defined in \cref{orthos}. For this purpose we rewrite it when $\mathcal{V}\rightarrow\mathcal{V}U$, where $U$ is an unitary matrix, maintaining
\begin{equation}
\left(\mathcal{V}U\right)\cdot \left(U^\dagger\mathcal{V}^\dagger\right)=\mathbb{1}.
\label{orthosa}
\end{equation}
This reshuffle permits us to concentrate the influence on the orthogonality relation of the first two rows for a general pair in the first row $(V_{14}, \, V_{15}) $ into just $(V'_{14}, \, 0)$. Thus with $\mathcal{V}\rightarrow\mathcal{V}U$, we now have that
\begin{equation}
       \mathcal{V}U=\begin{pmatrix}
V_{11}  &V_{12}  &V_{13} &V'_{14}&0  \\ 
V_{21}  &V_{22}  &V_{23} &V'_{24}&V'_{25} \\
.. &..  &.. &..&.. \\
..  &.. &..&..&.. \\
..  &..  &.. &..&.. \\
    \end{pmatrix}.
\end{equation}
The point is that, it is possible to choose $U$ such that is only affects the 4th and 5th columns of $\mathcal{V}$, thus leaving the usual $3\times 3 $ CKM mixing part unchanged.
So, the orthogonality relation of the first two rows is now reduced to only $4$ elements
\begin{equation}
V_{11}V^*_{21}  + V_{12}  V^*_{22}+ V_{13}  V^*_{23} +V'_{14} V'^*_{24} =0.
\label{four}
\end{equation}

In this way and by applying the same trick consecutively to all rows and columns\footnote{
For the the orthogonality of the columns, we rearrange the relation  $\mathcal{V}^\dagger\cdot \mathcal{V}=\mathbb{1}$ into $\left(\mathcal{V}^\dagger U^\dagger \right)\cdot \left(U\mathcal{V}\right)=\mathbb{1}$.},
we may continue to use the auxiliary picture displayed in \cref{betagamma} of a $4\times 4$ unitary matrix to express the influence of a much larger context on the orthogonality relations derived from rows and columns.

\section{Deviations from Unitarity with VLQ-singlets}

We proceed with our exploration of models containing vector-like quark (VLQ) iso-singlets, where the standard $3\times 3$ sub-matrix of $V_\text{CKM}$ is no longer unitary. 

While skipping many of the technical details regarding these models (as they have already been extensively described in \cite{Alves:2023ufm,Botella:2021uxz}), in the following, we focus on the properties of VLQ singlet models which may not only help explain the DUs of the first row of the $3\times 3$ CKM mixing, but may also (and this being main scope of this work) accommodate significant DUs in the second row, together with DUs in the columns.

\subsection{Embedding Non-Unitary CKM-Mixing \label{embed} }
Consider a model with $n_u$ up-type and $n_d$ down-type vector-like singlets. Then, the mixing matrix $V_{\text{CKM}}$ is a $(3+n_u)\times(3+n_d)$ non-unitary (and in general non-rectangular) matrix and may be written as \cite{Alves:2023ufm,Lavoura:1992qd,Branco:2021vhs}
\begin{equation}
 V_{\text{CKM}}=A^\dagger_u\, A_d,
 \label{ckm0}
\end{equation}
where $A_q$ ($q=u,d$) corresponds to the first $3$ rows of the unitary matrices $\mathcal{U}_q$ which diagonalize the $(3+n_q)\times(3+n_q)$ quark mass matrices $\mathcal{M}_{q}$, i.e.,
\begin{equation}
 \mathcal{U}^\dagger_L\ \mathcal{M}_q \ \mathcal{U}_R=   \mathcal{D}_q,
\end{equation}
with
\begin{equation}
 \mathcal{U}_q=
 \begin{pmatrix}
        A_q\\
        B_q
    \end{pmatrix},
    \label{uq}
\end{equation}
and where the block $B_{q}$ corresponds to the remaining $n_{q}$ rows of $\mathcal{U}_{q}$. 

The non-unitarity of $V_\text{CKM}$ leads to flavour changing neutral currents (FCNCs) at tree-level in the coupling of quarks to the $Z$ and higgs bosons. Their magnitude is controlled by the $(3+n_q)\times(3+n_q)$ matrices
\begin{equation}
 F_u = A^\dagger_u A_u =V_\text{CKM} V^\dagger_\text{CKM}, \quad \quad F_d = A^\dagger_d A_d =V^\dagger_\text{CKM} V_\text{CKM}.
\end{equation}

With respect to the condition of unitarity of $\mathcal{U}_q$ one has
\begin{equation}
   A_qA_q^\dagger=\mathbb{1}_{3\times 3} , \hspace{5mm}B_qB_q^\dagger=\mathbb{1}_{n_q\times n_q} , \hspace{5mm} A_qB_q^\dagger=0,
   \label{unit1}
\end{equation}
as well as
\begin{equation}
   A_q^\dagger A_q + B_q^\dagger B_q=\mathbb{1}_{(3+n_q)\times (3+n_q)}.
   \label{unit2}
\end{equation}

Although the $V_{\text{CKM}}$ mixing is no longer unitary (and in general non-square), one can still embed $V_{\text{CKM}}$ in a larger $(3+n_u+n_d)\times(3+n_u+n_d)$ auxiliary matrix 
\begin{equation}
 \mathcal{V}= 
 \begin{pmatrix}
        A^\dagger_u A_d & B_u^\dagger\\
        B_d & 0_{n_d\times n_u}
\end{pmatrix},
 \label{ckm1}
\end{equation}
 which is now unitary (and square) by construction \cite{Branco:1992wr,Alves:2023ufm}. In fact, its unitarity can easily be checked from \cref{unit1,unit2}. 

As we shall see, this matrix turns out to be a very useful tool when examining the physical consequences induced by the introduction VLQ singlets.

An important point in this construction is also the reverse implication. Suppose that we have some $(3+n_u+n_d)\times(3+n_u+n_d)$ parametrization of a unitary matrix $\mathcal{V}'$ which generically has the same characteristics as \cref{ckm1}, i.e., 
\begin{equation}
 \mathcal{V}'= 
 \begin{pmatrix}
        X & \hat{B}^\dagger\\
        \Tilde{B} & 0_{n_d\times n_u}
\end{pmatrix}.
\label{ckm2}
\end{equation}
Is it then feasible to pinpoint two unitary matrices $\mathcal{U}_{u}$ and $\mathcal{U}_{d}$ which will make it possible for us to identify $\mathcal{V}'$ with $\mathcal{V}$? Or equivalently: does the unitary matrix $\mathcal{V}'$ in \cref{ckm2} correspond to a model with $n_u$ up-type and $n_d$ down-type vector-like singlets, and will this effort permit us to constructively and comprehensibly explore the parameter-space of our model? 
Indeed, as we shall see, finding a suitable parametrization of a large auxiliary unitary, with special features as in \cref{ckm2} is crucial. 
 
\subsubsection*{Models with $n_u$ up and $n_d$ down-type VLQs: $(n_u U, n_d D)$}
In this work, we focus on the simplest models with the smallest number of VLQ-singlets which potentially could explain the current DUs. We denote these as $(n_u U, n_d D)$. Thus e.g., the $(2U,0D)$ case refers to the SM extended with just two up-type VLQ-singlets, while $(2U,1D)$ has two up-type and one down-type VLQ-singlets.
For each of these cases, we shall provide a method{\footnote{For the simplest cases of $(1U,0D)$ or $(0U,1D)$, this turns out to be a trivial exercise. }} in order to obtain the two unitary matrices $\mathcal{U}_{u}$ and $\mathcal{U}_{d}$ which lead to $\mathcal{V}$ as in eqs. (\ref{ckm1}, \ref{ckm2}). 

An important point in this notation is that by having e.g. only up-type VLQs does not exclude having down-type VLQs. If these have negligible mixing with other quarks, then effectively they can be neglected. Each time we refer to a notation with other type of quarks, we thus mean that these VLQs have meaningful mixing amongst each other (e.g. to account for deviations from unitarity in different rows).

It is also crucial to emphasize the role of the FCNCs effects to the validity of these models and their success in accommodating large DUs.
In the physical basis, the coupling of the quark neutral currents to $Z$ is proportional to $F^u=V_{\text{CKM}}V^\dagger_{\text{CKM}}$ for the up-quark fields and to $F^d=V^\dagger_{\text{CKM}} V_{\text{CKM}}$ for the down-quark fields \cite{Alves:2023ufm,Branco:2021vhs,Botella:2021uxz}. If the $V_{\text{CKM}}$ mixing is no longer unitary, this effect may be of great significance. This fact will then call for a comprehensive phenomenological analysis of these models.

\subsection{Deviations from Unitary with just one VLQ-singlet}
\label{onesinglet}

The possibility of having significant deviations from unitarity in the first row of the mixing has been widely studied in the context of extensions with vector-like quarks \cite{Alves:2023ufm,Branco:2021vhs,Botella:2021uxz}. Even though the FCNCs that emerge as a result of these deviations are constrained by a number of flavor observables, many of these models can still successfully accommodate all these restrictions. 

In the case of extensions with just one up-type iso-singlet, the FCNCs involve solely up-type quarks and are controlled by the current-matrix $F_u=V_{\text{CKM}}V^\dagger_{\text{CKM}}$. Some of the most stringent constraints to these currents come from $D^0-\bar{D}^0$ mixing, which now receives FCNC induced by tree-level contributions. 
With respect to this model, current experimental limits on the mixing parameter $x_D\equiv\Delta m_D/\Gamma_D$ impose \cite{Golowich:2007ka,ParticleDataGroup:2024cfk}
\begin{equation}\label{eq:F12}
    |F^u_{uc}|\lesssim 1.6\times 10^{-4},
\end{equation}
or, in terms of the elements of $\mathcal{V}$ (being a $4 \times 4$ matrix),
\begin{equation} \label{eq:V14V24}
    \left|\mathcal{V}_{14}\mathcal{V}^*_{24}\right|\lesssim 1.6\times 10^{-4}.
\end{equation}
Simultaneously, when one has just one up-type singlet, the following equality is valid:
\begin{equation}\label{eq:fcnc_1VLQ}
    |F^u_{uc}|^2 =(1-F^u_{uu})(1-F^u_{cc}),
\end{equation}
where we have
\begin{equation}\label{eq:fcnc_1VLQ_2}
    1-F^u_{uu}=\Delta_1^2, \quad \quad 1-F^u_{cc}=\Delta_2^2.
\end{equation}
Hence, by inducing DUs capable of addressing the CAA, i.e. $\Delta_1 \approx 0.04$, combining \cref{eq:F12,eq:fcnc_1VLQ} implies
\begin{equation}
  \Delta_2 \lesssim 4.1 \times 10^{-3}. 
\end{equation}
This, however, is not consistent with DUs of the second row where $\Delta_2 $ would minimally be of the order of first row $\Delta_1 \sim O(\lambda^2)$, let alone with the results recently proposed in \cite{Bolognani:2024cmr}, which, as shown in \cref{eq:Delta_2}, would require an even bigger $\Delta_2 \sim O(\lambda)$, where $\lambda$ stands for the Cabibbo angle.

In the case of models with more than one up-type VLQ, the constraints from FCNC can be relaxed and one has more freedom \cite{delAguila:1998tp}, as the relation in \cref{eq:fcnc_1VLQ} becomes an inequality
\begin{equation}
   |F^u_{uc}|^2 \leq (1-F^u_{uu})(1-F^u_{cc}).
\end{equation}
In these cases, there is no explicit incompatibility with the constraint of \cref{eq:F12}. 

In the case of extensions with one VLQ down-type singlet $B$, the most stringent constraint is also set by $D^0-\overline{D^0}$ mixing. In this context, new box diagrams are induced where the VLQ is allowed to run inside the loops along with the other SM down-type quarks. However, given the current experimental bound on the mass of such a field ($m_B>1.15$ TeV \cite{ATLAS:2024zlo}), and how it compares to the masses of down-type SM quarks, one expects the NP contribution involving solely the VLQ to completely dominate. With this in mind, we have \cite{Golowich:2007ka}
\begin{equation}
\left|V_{uB}V^*_{cB}\right|<2.8\times 10^{-4},
\end{equation}
which is slightly less stringent than \cref{eq:V14V24}. Nonetheless, similar conclusions apply. For instance, if $\Delta_1 \approx 0.04$ one would obtain in this case $\Delta_2 \lesssim 7 \times 10^{-3}$, which is incompatible with the DUs suggested in \cite{Bolognani:2024cmr}.

In conclusion, for the one VLQ-singlet case, it is not possible to find any region in parameter-space where the most significant flavor observables are within the limits imposed by experiment, while trying to accommodate the expected DU for the first row ($\Delta_1\approx 0.04$) and significant DUs in the second row, which appears in this case to be bounded by $\Delta_2\lesssim 7\times 10^{-3}$.

Motivated by this, in the following, we will explore scenarios with more than one VLQ.
If, however, more VLQs are introduced, more terms will be relevant, potentially leading to cancellations which will allow not only for large DUs, but also ensure that these do not generate too large contributions to flavor observables. 
As we shall see, a more careful numerical analysis performed in this work will confirm this.

\section{Deviations of Unitarity in the 2nd Row} 
Given the results of the previous section, we are encouraged to explore extensions of the SM with more than one VLQ isosinglet. 
It is our aim here to probe (in an almost systematic way), to what extent as diverse singlet VLQs can explain the deviations of unitarity of the first row, and above all, explain the DUs in the second row in conjunction with DUs in the columns.
\subsection{A Quasi-Systematic approach: Number of VLQs}

We first focus on a small model with just two VLQ singlets, where one is an up-type and the other a down-type VLQ-singlet: the $(1U,1D)$ case.
Although, as we shall see, not particular successful in accommodating large DUs in the second row, this is still an important case to study. Due to its reduced complexity, its simpler structure and parametrization, and subsequent relatively clear analysis, it can easily be generalized, which will help unravel other (subsequent) more complicated cases.

\subsection*{Two VLQ-singlets: the $(1U,1D)$ case}
In $(1U,1D)$ models, the mixing is represented by a $4\times 4$ non-unitary matrix. 
Now, following the outline in subsection \ref{embed}, it is useful to embed this $4\times 4$ matrix in an  auxiliary larger $5\times 5$ unitary matrix $\mathcal{V}$, which by construction has a matrix-element $\mathcal{V}_{55}=0$. 
Then, due to the larger structure of this mixing, one has the possibility of choosing from an extensive array of possible parameterizations for $\mathcal{V}$. Unfortunately, the majority of these are not particularly useful. 

In addition, faced by the restrictions imposed by experimental results,
it is crucial to choose a suitable parametrization which facilitates the search of regions of parameter-space where all relevant flavor observables are within their experimental limits.
In certain model-building cases, it may also be important to construct two unitary matrices $\mathcal{U}_{u}$ and $\mathcal{U}_{d}$ leading to $\mathcal{V}$. 
We recall that relating the $5\times 5$ auxiliary unitary $\mathcal{V}$ to two unitary matrices $\mathcal{U}_{u}$ and $\mathcal{U}_{d}$ not only proves that (the initially abstract) $\mathcal{V}$ precisely arises from a $(1U,1D)$ model, but, as we shall see, may also become relevant in order to coherently explore the parameter-space.

In view of all of this, we focus on the following specific parametrization structure for $\mathcal{V}$ 
\begin{equation}
\mathcal{V}= P_{45}\cdot (O_{23} K_3 O_{13}' O_{12})\cdot K_{23}' \cdot(O_{35} O_{25} O_{15})\cdot K_{23} \cdot (O_{34} O_{24} O_{13}),
\label{v1}
\end{equation}
where $P_{45}$ is the $(45)$-permutation and $O_{ij}$ are the simple orthogonal real rotations in the $ij$-plane, i.e.,
\begin{equation}
 P_{45} = \begin{pmatrix}
    1 & 0 & 0 & 0 & 0
    \\
    0 & 1 & 0 & 0 & 0
    \\
    0 & 0 & 1 & 0 & 0
    \\
    0 & 0 & 0 & 0 & 1
    \\
    0 & 0 & 0 & 1 & 0
 \end{pmatrix},\ 
 \left( O_{ij} \right)_{\alpha \beta} \equiv 
 \begin{cases}
    \sin \theta_{ij}&  \alpha = i, \beta = j,
    \\
    - \sin \theta_{ij}&  \alpha = j, \beta = i,
    \\
    \cos \theta_{ij} &  \alpha = \beta = i \text{ or } \alpha = \beta = j,
    \\
    \delta_{\alpha \beta}, & \text{otherwise},
\end{cases}
\end{equation}
while the $K$ matrices are diagonal unitary matrices with complex phases. The notation is self-evident:  $K_{3}=\diag(1,1,e^{i\delta},1,1)$ while $K_{23}=\diag(1,e^{i\delta_2},e^{i\delta_3},1,1)$ and $K'_{23}=\diag(1,e^{i\delta'_2},e^{i\delta'_3},1,1)$. This parametrization has a set of $14$ parameters: 9 real angles in the orthogonal rotations $O_{ij}$ and 5 complex phases in the $K$ phase matrices. 

Although, in general terms already preformed in the recent VLQ iso-singlets review \cite{Alves:2023ufm}, it is here also worthwhile to derive the number of physical parameters, by considering a minimal weak-basis for the mass matrices of the $(1U,1D)$ case, e.g. where
\begin{equation}
\mathcal{M}_u = 
\begin{pmatrix}
    r & c & c &  r
    \\
    0 & r & c & c 
    \\
    0 & 0 & r & c
    \\
    0 & 0 & 0 & r
 \end{pmatrix}\ ,\ \ \ 
\mathcal{M}_d = 
\begin{pmatrix}
    r & 0 & 0 & r 
    \\
    0 & r & 0 & r 
    \\
    0 & 0 & r & r 
    \\
    0 & 0 & 0 & r 
\end{pmatrix}
\end{equation}
Here, the $r$'s stand for real entries, while the $c$'s represent complex elements. Note that for this weak-basis, it is possible to have an extra real element, e.g. the element $(\mathcal{M}_u)_{14}$, using a rephasing of the VLQ left-handed fields $U_{L,R}$. This rephasing does not contribute to the mixing. When counting all parameters in the $\mathcal{M}_{u,d}$ and comparing with \cref{v1}, one has to keep in mind that the $8$ quark masses present in this model.

With regard to two unitary matrices which will then lead to $\mathcal{V}$, the parameterization proposed in \cref{v1} pays off. In this case, we can almost directly read $\mathcal{U}_{u}$ and $\mathcal{U}_{d}$ from its expression, choosing for instance
\begin{equation}
\mathcal{U}_{d}= O_{34} O_{24} O_{13},
\label{d1}
\end{equation} while for the hermitian conjugate of $\mathcal{U}_{u}$, we may take
\begin{equation}
 \begin{matrix}\mathcal{U}_{u}^\dagger&=P_{45}\cdot (O_{23} K_3 O_{13}' O_{12})\cdot K_{23}' \cdot(O_{35} O_{25} O_{15})\cdot K_{23}\cdot P_{45}\\ \\
&=(O_{23} K_3 O_{13}' O_{12})\cdot K_{23}' \cdot P_{45}\cdot(O_{35} O_{25} O_{15})\cdot P_{45}\cdot K_{23}\\ \\
&=(O_{23} K_3 O_{13}' O_{12})\cdot K_{23}' \cdot(\hat{O}_{34} \hat{O}_{24} \hat{O}_{14})\cdot K_{23},
\end{matrix}
\label{u1}
\end{equation}
where the hat notation serves to distinguish these $\hat{O}_{i4}$ from those ${O}_{i4}$ in \cref{v1,d1}.

Nevertheless, we note explicitly that this is merely a possible (and in this case, indeed useful) choice. This is because the matrices $\mathcal{U}_{u}^\dagger$ and $\mathcal{U}_{d}$ are not uniquely defined. Indeed, for any arbitrary unitary matrix $V_3$ with effective elements only in its upper-left $3\times 3$ sub-matrix, the matrices
$\mathcal{U}_{u}^\dagger\cdot V^\dagger_3$ and $V_3\cdot\mathcal{U}_{d}$ will lead to the same $\mathcal{V}$ as in \cref{v1}. All in all, in this construction, it is important to notice that, as said, $\mathcal{V}$ is a $5\times 5$ matrix with $\mathcal{V}_{55}=0$, while $\mathcal{U}_{u}^\dagger$ and $\mathcal{U}_{d}$ are effectively $4\times 4$ matrices. Although $\mathcal{V}$ is not simply obtained through $\mathcal{U}_{u}^\dagger\cdot\mathcal{U}_{d}$, nevertheless with the construction described here, each part of $\mathcal{U}_{u}^\dagger$ and $\mathcal{U}_{d}$, i.e., $A_u^\dagger$, $B_u^\dagger$ and $ A_d$, $B_d$, as mentioned in \cref{uq,ckm1}, falls into place in $\mathcal{V}$.

\subsubsection*{Physical Output of the $(1U,1D)$ case}
The parameter-space description formulated in terms of $\mathcal{V}$ given in \cref{v1} allows for a coherent analysis of the physical implications of our model with one up-type VLQ together with one down-type VLQ, in particular the evaluation of flavor observables.

At first glance this scenario appears to be much more promising in generating large DUs, particularly given the fact that it avoids the constraints in \cref{eq:fcnc_1VLQ}. Also, by working with a larger matrix $\mathcal{V}$, it is  in this way possible to choose a parametrization that allows for a distribution of the DUs, of the first and second rows, into different column-positions. This is very useful, since it, with this construction, it is more easy to evade regions of the parameter space excluded by dangerous FCNCs effects.

More precisely, large values for the DUs in the first and second row
\begin{equation}
   \Delta^2_{1}= |\mathcal{V}_{14}|^2+|\mathcal{V}_{15}|^2, \quad \quad \Delta^2_{2}= |\mathcal{V}_{24}|^2+|\mathcal{V}_{25}|^2, 
\end{equation}
can now be easily incorporated while suppressing both the FCNC coupling $F^u_{uc}$, which with the use of our auxiliary matrix $\mathcal{V}$ corresponds to $F^u_{uc}=-\mathcal{V}_{15}\mathcal{V}^*_{25}$, and the combination $\lambda^D_B\equiv V^*_{uB}V_{cB}=\mathcal{V}^*_{14}\mathcal{V}_{24}$, which (as discussed in \cref{onesinglet}) are the most relevant mixing combinations controlling the NP contributions to $D^0-\overline{D^0}$.

Additionally, one can check that although constraints such as \cref{eq:fcnc_1VLQ} now apply to both sectors, the equalities in \cref{eq:fcnc_1VLQ_2} no longer hold in general, as we have instead
\begin{equation}
    1-F^u_{uu}\geq \Delta^2_1, \quad \quad 1-F^u_{cc}\geq \Delta^2_2,
\end{equation}
meaning that many of the restrictions present in the one-singlet scenarios disappear.

\subsubsection*{Phenomenological Effects of Introducing Up and Down-Type VLQs}

The phenomenology of models with one VLQ singlet has been extensively studied in the past \cite{Alves:2023ufm} and, in many cases, the expressions for flavor observables in these scenarios can easily be generalized to the $(n_u U,0)$ or $(0,n_d D)$ cases with $n_{u,d}>1$. 
However, when extending the SM with singlets VLQs in both of the quark sectors, these generalizations are no longer valid. Entirely new diagrams arise, which are absent in either the separate $(n_u U,0)$ or the $(0,n_d D)$ scenarios. To our knowledge, the phenomenology of this mingling of quark sectors in VLQ singlet extensions of the SM has yet to be adequately addressed in the literature.

Here, we examine these new effects. We find that new important contributions do arise when both up- and down-type VLQs are considered. In particular, neutral meson mixings now receive additional contributions due to new penguin-loop diagrams. As an example, consider the diagrams of \cref{fig:exotic_diagrams} which contribute to $K^0-\overline{K^0}$ mixing.
\begin{figure}[h]
    \centering
\begin{tikzpicture}
  \begin{feynman}
    \vertex (q1) at (0, 1.5) {\(d\)};
    \vertex (v1) at (2.6,0); 
    \vertex (q3) at (7., 1.5) {\(s\)};
    \vertex (q2) at (0,-1.5) {\(s\)};
    \vertex (q4) at (7.,-1.5) {\(d\)};
    \vertex (v2) at (5.2, 0);
    \vertex (l1) at (0.86,1);
    \vertex (l2) at (0.86,-1);

    \diagram*{
      (q1) -- [fermion] (l1) -- [fermion, edge label=\(\alpha\)](v1) -- [fermion, edge label=\(\beta\)] (l2) -- [fermion] (q2);
      (q3) -- [fermion] (v2) -- [fermion] (q4);
      (v1) -- [boson, edge label=\(Z\)] (v2);
      (l1) -- [boson, edge label=\(W\)] (l2);
    };
    \draw[fill=red] (2.6,0) circle (3pt);
    \draw[fill=blue] (5.2,0) circle (3pt);
    \draw[fill=green] (0.86,1) circle (3pt);
    \draw[fill=green] (0.86,-1) circle (3pt);
  \end{feynman}
\end{tikzpicture}
\caption{Diagrams contributing to $K^0-\overline{K^0}$ mixing that include effects of mixing both with up- and down-type VLQ-singlets. Here, $\alpha,\beta$ represent up-type quarks. In green we highlight CKM couplings, while in blue(red) we highlight down(up)-type FCNC couplings.}
    \label{fig:exotic_diagrams}
\end{figure}

Apart from the aforementioned significant effects to $D^0-\overline{D^0}$ oscillations, important NP contributions can arise also in other neutral meson sectors. Notably, NP effects in the kaon sector are now enhanced both by loop contributions and will depend crucially on the mixing combination $\lambda^K_T\equiv V^*_{Ts}V_{Td}$, as well as the FCNC coupling $F^d_{sd}$, which with the use of our auxiliary matrix $\mathcal{V}$ corresponds to $F^d_{sd}=-\mathcal{V}^*_{52}\mathcal{V}_{51}$. 

The parameter $\epsilon_K$, measuring indirect CP violation in the neutral kaon sector, is probably the most sensitive to these NP effects. In general, it is given  by 
\begin{equation}
\begin{split}
    \epsilon_K =\left|\frac{\kappa_\epsilon\text{Im}\left[\left(\lambda^K_u\right)^{*2} M^K_{12}\right]}{\sqrt{2}\left(\lambda^K_u\right)^2 \Delta m_K }\right|,
\end{split} 
\label{epk}
\end{equation}
where $\lambda^K_i=V^*_{is}V_{id}$, $\kappa_\epsilon =0.92 \pm 0.02$ \cite{Buras:2008nn} is an effective suppression factor and $M^K_{12}$ is the off-diagonal element of the $K^0-\overline{K^0}$ mass matrix, which now incorporates contributions from penguin-diagrams coming from the mixing of VLQ in both quark sectors as represented in \cref{fig:exotic_diagrams}.

The relevant expressions for the most important flavor observables are presented in \cref{app:pheno}, where these implications are discussed in more detail.

To test the feasibility of this and subsequent models, it is necessary to include the experimental restrictions set by various observables. This includes not only the moduli of the mixing matrix elements $V_{ud}$, $V_{us}$, $V_{ub}$, $V_{cd}$, $V_{cb}$, which are determined from tree-level processes (not including $V_{cs}$), and the usual unitary triangle angles $\gamma$, $\beta$, $\beta_s$, the aforementioned $x_D$ and $\epsilon_K$ parameters, as well as other quantities such as the direct CP violation parameter $\epsilon'/\epsilon$, the branching ratio of the golden mode $K^+\to \pi^+ \nu\bar{\nu}$ and the $B^0_{d,s}-\bar{B}^0_{d,s}$ and $K^0-\bar{K}^0$ mixing parameters $\Delta m_{B_{d,s}}$ and $\Delta m_{K}$. All these observables were incorporated into a $\chi^2$ analysis. Here, we also imposed the current experimental bound on the mass of VLQ iso-singlets, $m_Q>1.15$ TeV \cite{ATLAS:2024zlo}.

For the $(1U,1D)$ case, we find that, were it not for the extra terms coming from the combinations of the two sectors, it would not be possible to conciliate a value for $x_D\lesssim 0.41 \%$ together with an acceptable value for $\epsilon_K$.
Fortunately, and also with the help of the contributions from the extra terms it is possible to find a region at $99.7\%$ CL in parameter-space where one obtains DUs for both the first and second row which are near $\Delta_i\approx 0.04$. Although not sufficient to reproduce the value of \cref{eq:Delta_2}, this is still a noteworthy result that may turn out to be important, depending on the evolution of the experimental determination of $|V_{cs}|$. 

Nevertheless as we shall see, a worthwhile outcome of the $(1U,1D)$ case with respect to our construction of the parametrization of the auxiliary unitary matrix $\mathcal{V}$ in \cref{v1}, is that it can easily be generalized and it will greatly help analyse and understand many features of other (subsequent more complicated) cases. 

For completion, we mention the two other cases with just two VLQs: the case $(2U,0D)$ with only two up-type VLQs, and the case $(0U,2D)$ with just two down-type VLQs.

\subsection*{Two up-type VLQs: the $(2U,0D)$ case}

For the $(2U,0D)$ case, the mixing matrix $V_{\text{CKM}}$ is a $5\times 3$ non-unitary matrix. As stated, here the previous $(1U,1D)$ case given in \cref{v1} proves itself to be very useful when constructing the appropriate auxiliary unitary matrix $\mathcal{V}$. Its parametrization can also be used in the case $(2U,0D)$ with just two up-type VLQs\footnote{For the $(2U,0D)$ model the mixing is a $5 \times 3$ matrix, so we may use a general $5 \times 5$ unitary matrix and rotate the two last columns as to obtain a $0$ in the $55$ position. This accounts for the parametrization of the $(1U,1D)$ model.}. However, the mixing $V_{\text{CKM}}$ here is the $5\times 3$ left part of $\mathcal{V}$.

Having only up-type VLQ singlet has its advantages. It avoids the occurrence of dangerous FCNC interactions in the down sector which would then contribute at tree-level to $B^0_{d,s}$ and $K^0$ meson mixings, and therefore to $\epsilon_K$. As for $x_D$, we may now avoid the potentially significant box-diagram contribution present in extensions with down-type singlets.

As before, we test this model by including the experimental restrictions set by various flavour observables. With regard to this, our conscientious choice of parametrization of $\mathcal{V}$, from \cref{v1}, turned out to again be very useful, and we easily find a region in parameter-space, where we can conciliate all necessary restrictions on the observables. This region \footnote{We also considered this region for the $(1U,1D)$ case.} is projected out onto a relatively simple and well-defined neighbourhood where
\begin{equation}
\begin{array}{ll}
O'_{13}=O_{35}=O_{25}=O_{34}=\mathbb{1},
\\
K_{23}= K_{23}'=\mathbb{1}.
\end{array}
\label{neib}
\end{equation}
Roughly speaking, this neighbourhood also corresponds to a Lagrangian of a $(2U,0D)$ model, where in the weak basis that the $3\times 3$ down-quark mass matrix $M_d$ is diagonal, the $5\times 5$ up-quark mass matrix $\mathcal{M}_u$ assumes the form
\begin{equation}
\mathcal{M}_u=
\begin{pmatrix}
  r&r&r&0&r\\  
  0&r&c&r&0\\  
  0&0&r&0&0\\  
  0&0&0&r&0\\  
  0&0&0&0&r\\  
\end{pmatrix},
\end{equation}
with the $r$'s standing again for real entries, while the $c$ is complex.
 
With regard to unitarity deviations, one finds that, at $99.7\% CL$, the DUs of the second row $\Delta_2$ can exceed the DUs of the first row and reach a maximum of $\Delta_2\approx 0.07$.

\subsection*{Two down-type VLQs: the $(0U,2D)$ case}

We briefly mention the mirror case of $(2U,0D)$, denoted by $(0U,2D)$ which by contrast has only two down VLQ-singlets. For convenience in this case, we choose a different approach with regard to the parametrization of parameter-space. This has to do with flavor observables such as $x_D$ which now receives dangerous contributions not only from $\mathcal{V}_{24}\mathcal{V}_{14}^*$, but also from $\mathcal{V}_{25}\mathcal{V}_{15}^*$, while the quantity $\epsilon_K$ depends crucially on $|F^d_{ds}|^2$.

For this $(0U,2D)$ case, we find a suitable parametrization with
\begin{equation}
\mathcal{V}= (O_{23} O_{34} O_{14})\cdot K_{23}' \cdot(O_{35} O_{25} O_{15})\cdot K_{23} \cdot (O_{23}' K_3 O_{13} O_{12}) ,
\label{v02}
\end{equation}
where the CKM mixing is the upper $3\times 5$ of this unitary matrix.
With regard to the results for this case, they roughly mimic the previous $(2U,0D)$ case.

\section{Towards Larger Deviations of Unitarity}
\label{chapter21}
Up until now, we have succeeded in generating deviations from unitarity considering only extensions of the SM very limited in the number of VLQs. As a result, we find that our outcome appears to be moderate as the deviations from unitarity obtained for the second row $\Delta_2 $, roughly accompany the DUs of the first row $\Delta_1$ and might at most reach $\Delta_2\approx 0.07$. 
In view of these conclusions, we now embark on an analysis of various models containing more than just two VLQs.

Next, we focus on models with more complex combinations of up and down singlet VLQs: $(2U,1D)$, $(2U,2D)$, $(3U,1D)$ and $(3U,2D)$.
This will add significantly more NP contributions to the flavor observables and increase the complexity of the analysis of the parameter-space. However, as
it turns out, it will also open the possibility for larger DUs in the second row, approaching the large DUs implied by the result of \cite{Bolognani:2024cmr}.

\subsubsection*{Extended Parameter-Spaces}
Apart from extra NP contributions to flavor observables, larger combinations in $(n_u U,n_d D)$ models induce larger parameter-spaces. For example, in $(2U,1D)$ models, one already has a $5 \times 4$ CKM mixing matrix. As stated previously, in these cases, it is useful to incorporate the mixing into a larger auxiliary unitary matrix $\mathcal{V}$. Nevertheless, even for the "smaller" $(2U,1D)$ case, this already results in a $6\times 6$ matrix. 
 
In addition to the increase in parameters, there is then the matter of choice of the parametrization of $\mathcal{V}$. 
As was emphasized before, this choice is crucial in order to comprehensively probe parameter-space and implications on flavor observables. 

Based on previous analysis for the $(1U,1D)$ case, we consider, e.g. for the "smaller" $(2U,1D)$ case, a direct generalization of \cref{v1}, choosing for $\mathcal{V}$   
\begin{equation}
\mathcal{V}= P_{46}\cdot (O_{23} K_3 O_{13}' O_{12})\cdot K_{23}''\cdot (O_{36} O_{26} O_{16})\cdot K_{23}' \cdot(O_{35} O_{25} O_{15})\cdot K_{23} \cdot (O_{34} O_{24} O_{13}),
\label{v6}
\end{equation}
which now has $7$ complex phases and $12$ angles. Here, $P_{46}$ stands for the $(46)$-permutation. 

For the even larger $(2U,2D)$, $(3U,1D)$ and $(3U,2D)$ cases, we make use of similar parametrizations. However, for these larger scenarios, we have to allow for even larger auxiliary unitary matrices: for the $(2U,2D)$ and $(3U,1D)$ cases a $7\times 7$ unitary matrix, while for the $(3U,2D)$ case, the $\mathcal{V}$ is $8\times 8$ matrix.  

With regard as to where one hopes to find suitable regions in parameter-space for all these cases, we follow similar steps as before. These were outlined in \cref{neib} for the useful $(1U,1D)$ case.

\subsection{Larger Deviations from Unitarity}

Now, exploiting these more complex scenarios and using the $\chi^2$ analysis outlined before, we were able to find $99.7\%$ CL regions in parameter space where these more populous models can accommodate larger DUs, compatible with the results in \cite{Bolognani:2024cmr}.

\begin{figure*}
\centering
\resizebox{0.6\textwidth}{!}{%
  \includegraphics[scale=1]{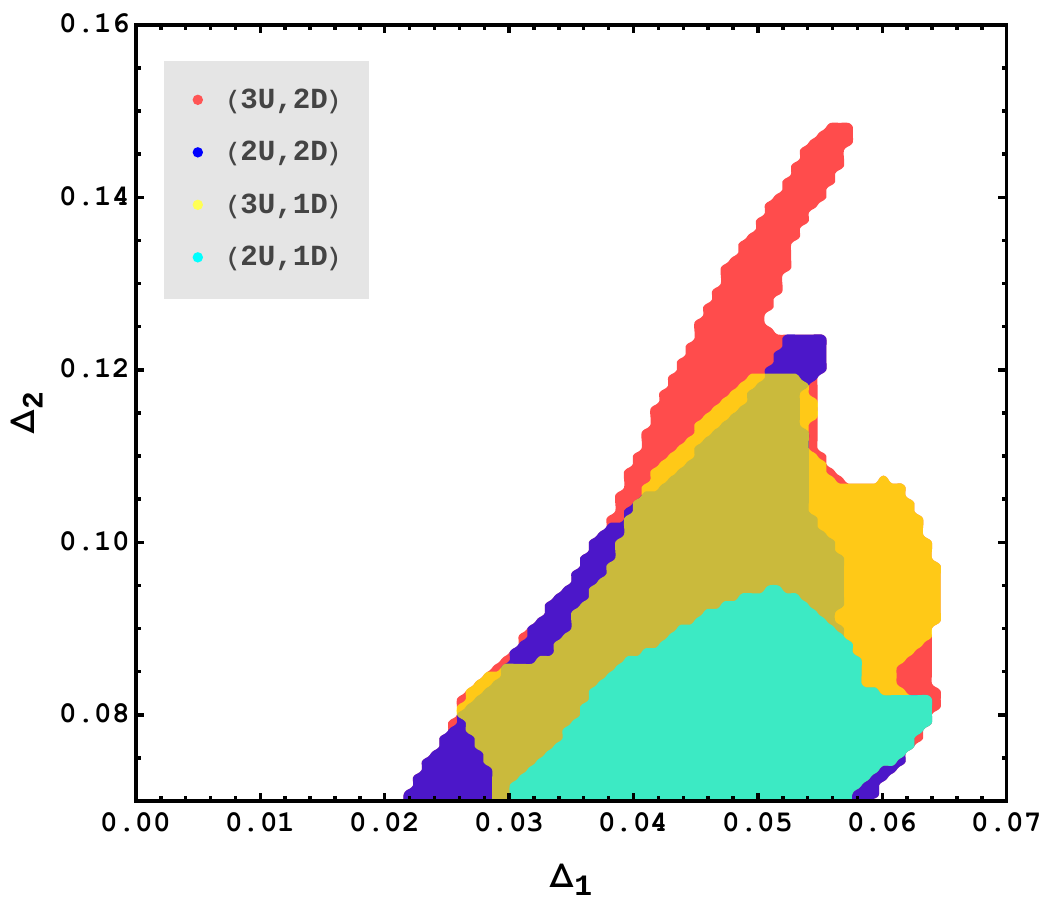}
}
\caption{Deviations from unitarity of the second row of the CKM matrix ($\Delta_2$) \textit{vs} deviations from unitarity of the first row of the CKM matrix ($\Delta_1$) for 4 distinct VLQ-singlet models. Each presented region corresponds to the 99.7\% CL for each model coming from the $\chi^2$ analysis. We focus on the $\Delta_2>0.07$ subregions.}	\label{fig:21}
\end{figure*}

We find that the deviations from unitarity in the second row increase with the increasing complexity of the models. In fact, we roughly obtain the following upper-limits for the DUs in the second row
\begin{equation}
\begin{matrix}
(2U,1D),  & \mathcal{V}\sim(6\times 6)&: & \Delta_2\lesssim 0.09 \\
(2U,2D),  & \mathcal{V}\sim(7\times 7)&: & \Delta_2\lesssim 0.12 \\ 
(3U,1D),  & \mathcal{V}\sim(7\times 7)&:  & \Delta_2\lesssim 0.12 \\ 
(3U,2D),  & \mathcal{V}\sim(8\times 8)&:  & \Delta_2\lesssim 0.15. \\
\end{matrix}
\end{equation}
In Fig. \ref{fig:21}, we provide additional details on these deviations from unitarity. There, we present the results for the region where $\Delta_2\geq 0.07$. We recall that this maximum for the DU of the second row was obtained for the case with just two VLQs. As can be seen from the plot we are able to achieve DUs in the second row, not far from the order of the Cabibbo angle. In fact, for the $(3U,2D)$ scenario, we even find $\Delta_2\lesssim 0.15\sim O(\lambda)$.

To complete our survey and as an example, we present a benchmark-point for the $(3U, 2D)$ case, where we obtain a $\Delta_1\simeq 0.057$ and a $\Delta_2\simeq 0.15$, corresponding to the following CKM matrix
\begin{equation}
|V_\text{CKM}|\simeq\begin{bmatrix}
0.97312 & 0.22307 & 0.00387 & 0.00550 & 0.02853 \\
 0.22400 & 0.96204 & 0.04086 & 0.03112 & 0.12315 \\
 0.00948 & 0.04396 & 0.94159 & 0.28405 & 0.00435 \\
 0.04807 & 0.06624 & 0.01716 & 0.00665 & 0.00859 \\
 0.01669 & 0.01344 & 0.12760 & 0.03825 & 0.00094 \\
 0.01325 & 0.04081 & 0.10663 & 0.03246 & 0.00529 
\end{bmatrix}\label{eq:benchmark}
\end{equation}
and VLQ masses $m_{T_1}\simeq 1.183$ TeV, $m_{T_2}\simeq 4.262$ TeV, $m_{T_3}\simeq 5.769$ TeV, $m_{B_1}\simeq 1.181$ TeV and $m_{B_2}\simeq 1.479$ TeV. This particular point sits within the $99.7\%$ CL region of \cref{fig:21} for a ($3U,2D$) model, with $\chi^2-\chi^2_\text{min}=5.68$, where $\chi^2_\text{min} \simeq 0.67$. 

We have thus demonstrated that, with the introduction of VLQs singlets, it is possible to achieve values for $|V_{cs}|$ closer to the result of \cite{Bolognani:2024cmr} ($\simeq 1.67\sigma$) and values of $\Delta_2$ around the expected value of \cref{eq:Delta_2} ($\simeq 1.81\sigma$).
Additionally, we note that the obtained mass spectrum includes many VLQs with $\sim 1$ TeV masses, which could be accessible to future experiments.

\begin{figure*}
\centering
\resizebox{\textwidth}{!}{%
  \includegraphics[scale=1]{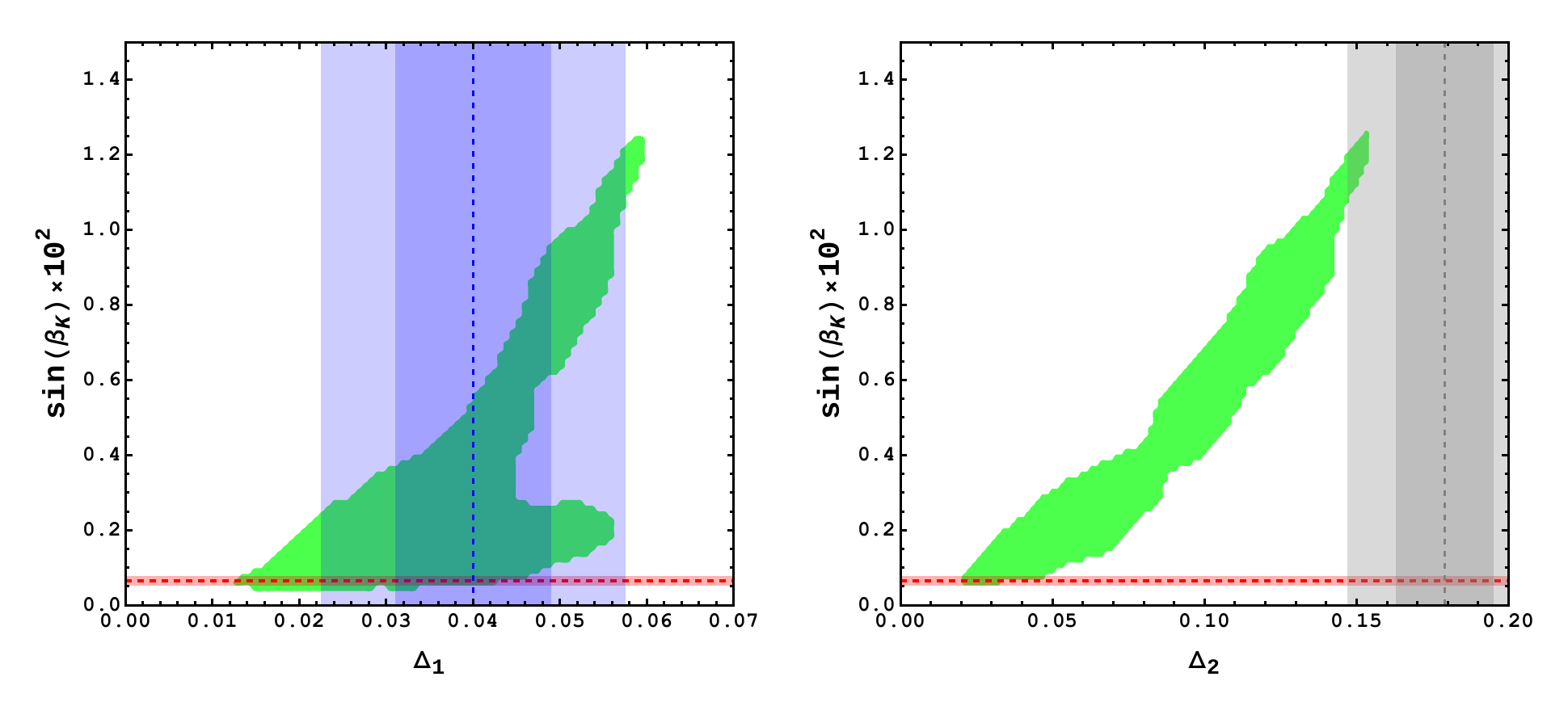}
}
\caption{For a (3$U$,2$D$) VLQ-singlet model, we present in green on the left (right) the allowed regions for $\sin \beta_K$ vs the deviation from unitarity of the first (second) row of the CKM matrix, $\Delta_1$ ($\Delta_2$), following a $\chi^2$ analysis together with a 99.75\% CL region. In red, we have the 3$\sigma$ region for $\sin\beta_K$ in the SM. As a guideline in blue, we depict the 1$\sigma$ and 2$\sigma$ regions for $\Delta_1$ coming from \cref{eq:Delta_1}, while in gray we do the same for $\Delta_2$ using the result of \cref{eq:Delta_2}.}	\label{fig:betaK}
\end{figure*}

Finally, we point out two additional important results. First, we note the outstanding effects of VLQ-singlet extensions on the rephasing invariant phase $\beta_K$. These are presented in \cref{fig:betaK} for the $(3U,2D)$ scenario, showing that huge enhancements up to $\sin \beta^{(3,2)}_K\sim 20 \sin \beta^\text{SM}_K$ are possible\footnote{Updating the upper bound for $\chi^\prime$ set in Section 2 of \cite{Aguilar-Saavedra:2004roc}, by using current PDG data, we now get $|\chi^\prime|\leq 0.03$ at $95\%$CL, which does not exclude any region in \cref{fig:betaK}.}. This result is particularly interesting in view of the analysis in \cite{Bediaga:2022sxw} (as aforementioned in \cref{ckmphase}) where it was suggested that the measured value for the CP asymmetries $\Delta A_\text{CP}$ may be accommodated with a doubling of $\sin{\beta_K}$ from its SM value.

Second, we find, that in the regions with large $\Delta_2$, the values of $|V_{tb}|$ may be significantly smaller than the PDG value. The benchmark point of \cref{eq:benchmark} resides in such a region, where remarkable large DUs also arise in the third row ($\Delta_3\sim 0.3$). The importance of this result will depend crucially on the evolution of future experimental determinations of $|V_{tb}|$. However, given the current large uncertainties in the measurements of this element of the third row, we expect, in contrast, the status of CKM unitarity in the second row to become clear sooner.

\section{Conclusions} \label{sec:conclusion}
	
Starting from the assumption that the CKM matrix is a $3 \times 3$ unitary matrix, we derived relevant relations between its elements, which include the moduli and its complex phases. We have found that not all of these are totally compatible with experiment and/or the assumption of the $3 \times 3$ unitarity. In particular, in the relations for the columns, we observe tensions between the current values of $V_{ud}$ and $V_{td}$, and the new values for $V_{cs}$ presented in \cite{Bolognani:2024cmr} and $V_{ts}$. We further note that a possible large experimental result for the CKM complex phase $\beta_K$ can signal deviations from unitary. We also point out that (with regard to future results and models with VLQ-singlets) inconsistencies between moduli and complex phases not matching experiment, may even render deviations from unitarity, exclusively attributed to physics beyond the SM solely induced by VLQ singlets, to be unsustainable.

With regard to the deviations from unitarity induced by VLQ singlets, we focused on models which may explain DUs of the first row but also induce significant DUs of the second row together with DUs in the columns. 
First, we made a thorough analysis of VLQ models with the lowest amount of singlets. We then went on to consider more complicated and elaborate models in order to obtain larger deviations from unitarity.

A key aspect of the analysis of all these models was establishing the appropriate parametrization for each case, which then became a fundamental piece in achieving coherent and systematic explorations of the parameter space.

We tested the feasibility of each model, including relevant constraints from a set of key observables, such as the moduli of $V_{ud}$, $V_{us}$, $V_{ub}$, $V_{cd}$, $V_{cb}$ and the unitary triangle angles $\gamma$, $\sin(2\beta)$ and $2\beta_s$. Besides these, special attention was given to flavor observables $x_D$ and $\epsilon_K$, as well as the CP violation parameter $\epsilon'/\epsilon$, the branching ratio of $K^+\to \pi^+ \nu\overline{\nu}$ and the $B^0_{d,s}-\overline{B^0}_{d,s}$ mixing parameters $\Delta m_{B_{d,s}}$. All these were inserted into a global $\chi^2$ test.

We concluded that scenarios with just one VLQ-singlet, while having the expected DU for the first row of $\Delta_1\approx 0.04$, are not capable of inducing sizable deviations in the second row. Similar conclusions apply in the case $(1U,1D)$ with one up-type and one down-type singlet VLQ.
Slight improvement is obtained for the case $(2U,0D)$ with only two up-type singlet VLQs, and the case $(0U,2D)$ with just two down-types. However, the DUs of the second row only exceed the DUs of the first row, on average, up to $21\%$, and cannot be larger than an upper-limit of $\Delta_2 \lesssim 0.07$.

Finally, it was possible to obtain notable results for the deviations from unitarity in the quark mixing and achieve very large DUs, even up to the order of the Cabibbo angle, in the second row of the CKM mixing. This was possible via the introduction of more VLQ singlets, thus expanding the parameter space. We focused on more complex VLQ singlet models, which throw in together a diverse number of up and down-type singlet-VLQs: the combinations $(2U,1D)$, $(2U,2D)$, $(3U,1D)$ and $(3U,2D)$.

Novel NP contributions generated by the mingling of the two sectors and the unique possibilities offered by the larger parameter space make it possible for these models to incorporate large deviations from unitarity in the second row, even as far as $\Delta_2\approx 0.15\sim O(\lambda)$. In fact, with such large DUs we can now accommodate the results of \cite{Bolognani:2024cmr} at $<2\sigma$.

We also point out that the new contributions from the mingling of the quark-sectors, presented here, have up until now been underestimated or even non-described in the literature.

Finally, we note that VLQ singlet models lead to large enhancements of the rephasing invariant phase $\beta_K$, up to 20 times of the SM prediction, and may lead to large DUs in the third row, directly connected to reduced values of $|V_{tb}|.$ These results emphasize the importance of precise determination of these quantities as probes for potential NP beyond the SM.

\section*{Acknowledgments}

We would like to thank Miguel Nebot for his advice on how to implement the parameter space scans performed in this work. We would also like to thank M. N. Rebelo for the many insights and careful reading of the paper.
This work was partially supported by Fundação para a Ciência e a Tecnologia (FCT, Portugal) through the projects CERN/FIS-PAR/0002/2021, 2024.02004.CERN and CFTP-FCT Unit UID/00777/2025 (https://doi.org/10.54499/UID/00777/2025), which are partially funded through POCTI (FEDER), COMPETE, QREN and EU.
FJB received support from \textit{Agencia Estatal de Investigaci\'on}-\textit{Ministerio de Ciencia e Innovaci\'on} (AEI-MICINN, Spain) under grants PID2022-139842NB-C22/AEI/10.13039/501100011033 (AEI/FEDER, UE) and from \textit{Generalitat Valenciana} under grant CIPROM 2022-36. J.F.B. acknowledges funding from
Fundação para a Ciência e a Tecnologia (FCT) through Grant PRT/BD/154581/2022 (https://doi.org/10.54499/PRT/BD/154581/2022). F.A. acknowledges funding from
Fundação para a Ciência e a Tecnologia (FCT) through Grant UI/BD/153763/2022 (https://doi.org/10.54499/UI/BD/153763/2022).

\newpage

\appendix

\section{Phenomenological Effects of $(n_u U, n_d D)$ models}
\label{app:pheno}

With regard to BSM physics induced by VLQ-singlets where one has a mingling of the up and down type quarks, it is (also) necessary to consider the new contributions generated by the respective combinations. 

In the kaon sector, the matrix element for neutral kaon-mixing includes the following terms
\begin{equation}
    M^K_{12}=\frac{G^2_F M^2_W m_K f^2_K B_K}{12\pi^2} \left(\Delta^K_\text{box}+\Delta^K_\text{tree}+\Delta^K_\text{scalar-box}+\Delta^K_\text{penguin}\right).
\end{equation}
The first term stems from the $W$-mediated box-diagram contributions to $K^0-\overline{K^0}$, which include the totality of the leading SM contributions, as well as the new box diagrams where the up-type VLQs run inside the loop. These contributions can be written as
\begin{equation}
   \Delta^K_\text{box}=\sum_{\alpha,\beta} \eta^K_{\alpha\beta} \lambda^K_\alpha\lambda^K_\beta S(x_\alpha,x_\beta),
\end{equation}
where $\alpha,\beta=c,t,T_1,\ldots,T_{n_u}$, $\lambda^K_\alpha=V^*_{\alpha s}V_{\alpha d}$ and $S(x_\alpha,x_\beta)$ with $x_\alpha=\left(m_\alpha/M_W\right)^2$ are the Inami-Lim box-loop functions \cite{Inami:1980fz,Buchalla:1990qz,Buchalla:1995vs}. We follow the procedure of \cite{Buchalla:1990qz,Alves:2023ufm} and use $\eta^K_{\alpha \beta}\simeq\eta^K_{tt}$ for the QCD corrections of the VLQ terms.

In turn, the presence of down-type VLQs will induce $Z$-boson mediated tree-level contributions to this mixing, given by
\begin{equation}
   \Delta^K_\text{tree}=\frac{4\pi s^2_W}{\alpha}\ \eta^K_Z (F^d_{sd})^2.
\end{equation}
Additionally, Higgs-mediated FCNCs will induce scalar-box contributions, leading to \footnote{This result is only known in the limit of large VLQ masses ($m^2_{Q}\gg m^2_{W,Z}$). This is the first time this expression has been presented in this particular form, i.e. making use of the $N_0$ function, usually appearing in the correction term of penguin-diagrams with VLQs. Still, it reproduces the previous results of \cite{Belfatto:2021jhf,Ishiwata:2015cga} in the appropriate limits.}
\begin{equation}
   \Delta^K_\text{scalar-box}\simeq 2 \eta^K_Z\sum_{\alpha,\beta} N_0(x_\alpha,x_\beta)\Lambda^K_\alpha \Lambda^K_\beta.
\end{equation}
Here $\Lambda^K_\alpha=F^{d^*}_{\alpha d}F^{d}_{\alpha s}$ and the relevant loop function is
\begin{equation}
    N_0(x_\alpha,x_\beta)\equiv \frac{x_\alpha x_\beta}{8(x_\alpha-x_\beta)}\log\left(\frac{x_\alpha}{x_\beta}\right).
\end{equation}
In this limit, we expect these contributions to be completely dominated by the diagrams involving only VLQs in the loop, and therefore it suffices to take $\alpha,\beta=T_1,\ldots,T_{n_u}$ in the sum. 

We also consider EW penguin-diagram contributions to $K^0-\overline{K^0}$. It is here where we find important new contributions that have been absent from the literature. In the presence of only down-type VLQs ($n_u=0$), the sum over all penguin-diagrams contributions is known to result in terms proportional to the Inami-Lim function $Y_0(x_i)$ \cite{Barenboim:1997pf}, leading to an overall contribution dominated by the top-quark term of the form $\sim F^d_{sd}\ \lambda^K_t Y_0(x_t)$. Then, one could naively assume that, with the inclusion of up-type VLQs (now also allowed to run inside the loops), generalizing this contribution would amount to having instead a contribution of the form  $\sim F^d_{sd}\sum_{\rho}\lambda^K_\rho Y_0(x_\rho)$ with $\rho=t,T_{1},\ldots,T_{n_u}$. This would be the case if we were instead in the presence of additional chiral generations of quarks beyond the three standard ones. However, the vector-like nature of these heavy-quarks induces up-type FCNCs at tree-level that modify a specific type of penguin-diagram contribution, namely the one displayed in \cref{fig:exotic_diagrams} where the up-type quarks in the loop couple to the $Z$-boson. This leads to a suppression of the overall penguin-loop term when compared to the straightforward scenario with extra chiral quark generations.

This suppression can be accounted for by a correction term, leading to the full form\footnote{Analogous terms are included for the matrix elements for the $B^0_{d,s}-\overline{B^0_{d,s}}$ mixings.}  
\begin{equation} \label{eq:penguin_term}
   \Delta^K_\text{penguin}\simeq -8 \eta^K_{tt}F^d_{sd}\left[\left( \sum_{\rho} \lambda^K_\rho Y_0(x_\rho)\right) + A_{ds}\right],
\end{equation}
where we take $\rho=t,T_1,\ldots,T_{n_u}$.
The correction term is proportional to a factor $A_{ds}$, given by \cite{Botella:2021uxz,Nardi:1995fq}
\begin{equation}\label{eq:Ads}
 A_{ds}=\sum_{\gamma,\delta}V_{\gamma d}(F^u-\mathbb{1})_{\delta \gamma }V^*_{\delta s}N_0(x_\gamma,x_\delta),
\end{equation}
with $\gamma,\delta=c,t,T_1,\ldots,T_{n_u}$. This correction is crucial in keeping the dependence of the penguin term on the VLQ masses in control, as it cancels the linear dependence in $x_T=m^2_T/M^2_W$ of $Y_0(x_T)$, resulting in an overall $\log x_T$ dependence and enforcing the expected decoupling behavior in the limit of large VLQ masses.

Similarly, the flavor observable $x_D$, can now be written as
\begin{equation}
    x_D=\frac{G^2_F M^2_Wf^2_DB_Dm_D}{6\pi^2\Gamma_D}\left|\Delta^D_\text{box}+\Delta^D_\text{tree}+\Delta^D_\text{scalar-box}+\Delta^D_\text{penguin}\right|,
\end{equation}
where the new term, analogously to that of \cref{eq:penguin_term}, can be written as (with the relevant Inami-Lim function now being $X_0$) 
\begin{equation}
   \Delta^D_\text{penguin}\simeq -8 F^u_{cu}\left[\left( \sum_{l} \lambda^D_l X_0(x_l)\right) + A_{cu}\right],
\end{equation}
with $l=B_1,\ldots,B_{n_d}$, while
\begin{equation}
    A_{cu}=\sum_{i,j}V_{ci}(F^u-\mathbb{1})_{ij}V^*_{ uj}N_0(x_i,x_j),
\end{equation}
with $i,j=s,b,B_1,\ldots,B_{n_d}$ and $\lambda^D_l=V^*_{ul}V_{cl}$.

The terms $\Delta^D_\text{box}$ and $\Delta^D_\text{scalar-box}$ are completely dominated by the VLQs terms and can be written as
\begin{equation}
\Delta^D_\text{box}=\sum_{i,j}\lambda^D_i\lambda^D_j S(x_i,x_j),
\end{equation}
with $i,j=B_1,\ldots,B_{n_d}$, and
\begin{equation}\label{eq:scalar_box}
   \Delta^D_\text{scalar-box}=2\eta^D_Z\sum_{\alpha,\beta} N_0(x_\alpha,x_\beta)\Lambda^D_\alpha \Lambda^D_\beta,
\end{equation}
with $\alpha,\beta=T_1,\ldots,T_{n_u}$ and $\Lambda^D_\alpha=F^{u^*}_{\alpha c}F^{u}_{\alpha u}$.
The remaining term is given by
\begin{equation}
   \Delta^K_\text{tree}=\frac{4\pi s^2_W}{\alpha}\ \eta^D_Z (F^u_{cu})^2.
\end{equation}

	\providecommand{\noopsort}[1]{}\providecommand{\singleletter}[1]{#1}
	
	\providecommand{\href}[2]{#2}\begingroup\raggedright\endgroup
		
\end{document}